\documentclass[journal]{IEEEtran}
\IEEEoverridecommandlockouts
\usepackage[sort&compress,numbers]{natbib}

\usepackage{makeidx,epsfig}
\usepackage{setspace,graphicx,multirow}
\usepackage{amsmath,amssymb,amsfonts}
\usepackage[small]{caption}
\usepackage{subcaption}
\usepackage{graphicx}
\usepackage{epstopdf}
\newtheorem{remark}{\bf Remark}

\usepackage{array}
\usepackage{url}
\usepackage{multirow}
\usepackage{hhline}
\renewcommand{\arraystretch}{1.5}

\usepackage{xcolor}
\usepackage{textcomp}

\usepackage{bm}
\usepackage{mathtools}
\usepackage{stfloats}
\usepackage[ruled,linesnumbered]{algorithm2e}
\usepackage{varwidth}

\makeatletter
\def\BState{\State\hskip-\ALG@thistlm}
\makeatother
\usepackage{fixltx2e}

\usepackage{enumitem}

\makeatletter
\newcommand\tinyv{\@setfontsize\tinyv{7pt}{9}}
\makeatother

\usepackage{authblk}

\usepackage{bbm}
\usepackage{color, soul, framed}

\ifodd 0
\newcommand{\rev}[1]{{\color{red}#1}} 
\newcommand{\com}[1]{\textbf{\color{blue} (COMMENT: #1)}} 

\else
\newcommand{\rev}[1]{#1}

\newcommand{\com}[1]{}
\fi

\usepackage{soul}

\usepackage{booktabs}
\usepackage{diagbox}

\begin{document}
\bibliographystyle{IEEEtran}
\bstctlcite{IEEEexample:BSTcontrol}

\title{Spatial Deep Learning for Site-Specific Movement Optimization of Aerial Base Stations}
\author{Jiangbin~Lyu, ~\textit{Member,~IEEE},
		Xu~Chen,~\textit{Student Member,~IEEE},
        Jiefeng~Zhang,~\textit{Student Member,~IEEE},
        Liqun~Fu, ~\textit{Senior Member,~IEEE}%
\thanks{Manuscript submitted to IEEE Trans. Wireless Communications on 15 Jan. 2023; revised 11 Sep. 2023; accepted 5 Dec. 2023. This work was supported in part by the Natural Science Foundation of Fujian 
	Province (No. 2023J01002), the Natural Science Fundation of Xiamen (No. 3502Z202372002), the Guangdong Basic and Applied Basic Research Foundation (No. 2023A1515030216), the National Natural Science Foundation of China (No. U23A20281, No. 61801408), and the Fundamental Research Funds for the Central Universities (No. 20720220078).}
\thanks{
The authors are with the School of Informatics, Xiamen University, China, and also with the Shenzhen Research Institute of Xiamen University, China (email: \{ljb, liqun\}@xmu.edu.cn; \{haxrd, zhangjiefeng\}@stu.xmu.edu.cn). \textit{Corresponding author: L. Fu}.}%
}%
\maketitle
\begin{abstract}
Unmanned aerial vehicles (UAVs) can be utilized as aerial base stations (ABSs) to provide wireless connectivity for ground users (GUs) in various emergency scenarios.
However, it is a NP-hard problem with exponential complexity in $M$ and $N$, in order to maximize the coverage rate of $M$ GUs by jointly placing $N$ ABSs with limited coverage range.
The problem is further complicated when the coverage range becomes irregular due to site-specific blockages (e.g., buildings) on the air-ground channel, and/or when the GUs are moving.
To address the above challenges, we study a multi-ABS movement optimization problem to maximize the average coverage rate of mobile GUs in a site-specific environment.
The Spatial Deep Learning with Multi-dimensional Archive of Phenotypic Elites (SDL-ME) algorithm is proposed to tackle this challenging problem by 
1) partitioning the complicated ABS movement problem into ABS placement sub-problems each spanning finite time horizon; 
2) using an encoder-decoder deep neural network (DNN) as the emulator to capture the spatial correlation of ABSs/GUs and thereby reducing the cost of interaction with the actual environment;
3) employing the emulator to speed up a quality-diversity search for the optimal placement solution;  and
4) proposing a planning-exploration-serving scheme for multi-ABS movement coordination.
In particular, the locations of ABSs/GUs are converted into grid pattern representations, whose dimension and associated DNN complexity are invariant with arbitrarily large $M$ and/or $N$.
Moreover, the virtual emulator-planning combined with the actual site-deployment effectively compensates for the prediction errors due to model approximation.
Numerical results demonstrate that the proposed approach significantly outperforms the benchmark Deep Reinforcement Learning (DRL)-based method and other two baselines in terms of average coverage rate, training time and/or sample efficiency.
Moreover, with one-time training, our proposed method can be applied in scenarios where the number of ABSs/GUs dynamically changes on site and/or with different/varying GU speeds, which is thus more robust and flexible compared with conventional DRL-based methods.

\end{abstract}


\section{Introduction}\label{sec:introduction}
With their high mobility and reducing cost, unmanned aerial vehicles (UAVs) have attracted increasing interests in military and civilian domains in recent years.
In particular, integrating UAVs into wireless communication networks as aerial base stations (ABSs) to assist terrestrial communication infrastructure in various emergency scenarios such as battlefields, disaster scenes and hotspot events, has been regarded as an important and promising technology \cite{ZengAccessing}.

One of the key problems in UAV-aided communication systems is to find applicable placement of $N$ ABSs with limited coverage range in order to achieve maximum coverage of $M$ (static) ground users (GUs) \cite{2017-Lyu-Placement, 2016-Galkin-Deployment, 2016-Mozaffari-Efficient}, which is known to be a \textit{NP-hard problem} with exponential complexity in $M$ and $N$ \cite{2017-Lyu-Placement}.
A tutorial in \cite{2021-Pham-PlacementIntro} discusses the ABS placement problem and the most commonly used schemes in scenarios with free space (FS) or non-FS propagation in recent literature.
In particular, some algorithms including the spiral algorithm \cite{2017-Lyu-Placement}, K-means algorithm \cite{2016-Galkin-Deployment}, circle packing theory \cite{2016-Mozaffari-Efficient}, and user-majority based adaptive UAV deployment \cite{8760267}, are proposed to solve the type of problems with dominant line-of-sight (LoS) or probabilistic LoS/non-LoS (NLoS) channel model \cite{2014-Hourani-Optimal}, under which each ABS has a \textit{uniform} coverage range.
However, due to \textit{site-specific blockages} (e.g., buildings), the above channel models might fail to capture the fine-grained structure of LoS or NLoS propagation at specific ABS and GU locations \cite{2019-Lyu-RadioMap, 2020-Qiu-Placement}.
For example, with a slight change of its position, an ABS might transit from LoS to NLoS propagation to the GU due to building edges.
This critically affects the ABS-GU channel and further complicates the problem.

Some efforts have been made using deep learning (DL) to learn \textit{site-specific channel} information \cite{2021-Jeffrey-Learning, 2019-Yu-Spatial}, and/or using radio map \cite{2019-Lyu-RadioMap} to construct/utilize spatial channel distribution \cite{LyuIoTJ3d,2019-Zhang-RadioMap-Path, 2021-Zeng-Navigation-RadioMap,2022-Romero-Placement-RadioMap}.
In \cite{2021-Jeffrey-Learning}, the authors propose to use an end-to-end neural network to learn a site-specific probing codebook in order to predict the optimal narrow beam for beam alignment.
The authors in \cite{2019-Yu-Spatial} propose a DL-based method for the optimal scheduling of interfering links in a dense wireless network with full frequency reuse.
The proposed methods in \cite{2021-Jeffrey-Learning, 2019-Yu-Spatial} are able to reduce the time/computational overhead of channel estimation and/or schedule links efficiently based on geographic locations of the devices, which yet are not directly applicable to the ABS placement/movement problems.
In the context of UAV communications,
radio map has been utilized to represent site-specific spatial distribution of average received power radiated from given transmitting source(s),  e.g., the fixed ground base stations for cellular-connected UAV \cite{LyuIoTJ3d, 2019-Zhang-RadioMap-Path, 2021-Zeng-Navigation-RadioMap}, or the ABSs \cite{2022-Romero-Placement-RadioMap} to provide ground coverage.
For ABS placement, the authors in \cite{2022-Romero-Placement-RadioMap} leverage on a given spatial loss field (SLF) function to construct the radio map with low complexity, whereas how to obtain/store site-specific SLF for any given ABS-GU location pair with high sample efficiency is yet to be addressed. 
Other authors in \cite{2021-Chen-UAV-Replay} partially circumvent this challenge by developing adaptive UAV positioning strategy with on-site LoS condition measurements for a given pair of UAV-relay and GU, whereas multi-ABS/multi-GU scenarios are yet to be considered.


The ABS placement problem can be further complicated due to \textit{GU mobility}, which brings additional complexity and the practical requirement of finding desired solutions within limited time.
In this regard, machine learning methods including DL and reinforcement learning (RL)/deep RL (DRL) have been developed/applied to solve 
complicated problems of UAV joint optimization considering multiple factors such as UAV trajectory design, user association, resource allocation and power consumption \cite{2019-Liu-UAV-MISC, 2019-Chen-UAV-MISC, 2021-Wang-UAV-MISC} (see the recent survey \cite{2020-Ullah-UAVs} for more references).
In particular, RL/DRL methods 
have recently been applied to tackle the multi-UAV movement optimization problems \cite{2018-Liu-Energy, 2020-Liu-Distributed, 2019-Liu-Reinforcement, 2019-Wang-Multi-UAV, 2021-Zhang-Three-Dimension, 2020-Lyu-Codesign}.
In terms of ABS coverage and energy consumption trade-off, a DRL-based approach is proposed in \cite{2018-Liu-Energy} to achieve energy-efficient and fair communication coverage. 
A decentralized DRL-based framework is further proposed in \cite{2020-Liu-Distributed} to provide energy-efficient long-term communication coverage.  
The authors in \cite{2019-Liu-Reinforcement}
propose a genetic algorithm-based K-means algorithm to partition GUs into clusters, and further apply the Q-learning algorithm in each cluster for ABS movement.
The authors in \cite{2019-Wang-Multi-UAV} and \cite{2021-Zhang-Three-Dimension} both formulate the ABS movement problem as a constrained markov decision process, and propose the dueling Deep Q-network (DQN) and/or constrained DQN algorithms to maximize the downlink capacity with full coverage constraint.
\rev{However, the above works typically assume uniform coverage range in a generic
environment, whereas the \textit{site-specific LoS/NLoS propagation} scenario is yet to be considered.

For \textit{site-specific multi-ABS movement} optimization, the authors in \cite{2020-Lyu-Codesign} propose a single-agent Deep Deterministic Policy Gradient (DDPG) based approach to maximize the average sum-rate of all GUs via UAV dynamic movement and communication co-design, whereas the tested network is relatively small (e.g., with $M=10$ and $N=2$).}
On the other hand, our early work in \cite{2020-Qiu-Placement} proposes a Double DQN with Prioritized Experience Replay (PER-DDQN) to address the site-specific ABS placement problem with a moderate network size (e.g., $M=80$ and $N=10$). However, a straightforward extension to ABS movement optimization encounters further difficulties.  
In particular, the action space of RL/DRL methods grows \textit{combinatorially} with $N$ and the number of steps to explore.
Moreover, a new neural network (NN) model in DRL would often need to be re-trained in order to cater for \textit{network changes} (e.g., ABSs/GUs turning on/off, varying GU speeds, etc.), which calls for more timely adaptation and more flexible design. 

To circumvent the above difficulties, we shift our mindset and attempt a different approach other than DRL.
First, the ABS movement problem is partitioned into a time series of ABS placement sub-problems, each of which aims for maximum GU coverage under given GU locations, and thus
amounts to a \textit{pattern matching/search problem}.
Second, a state-of-the-art search algorithm called Multi-dimensional Archive of Phenotypic Elites (MAP-Elites) \cite{2015-MAP-Elites} is adopted with tailored modifications to solve each ABS placement sub-problem.
Third, \rev{an environment emulator is built to predict the site-specific coverage status of all GUs in the actual environment and assist fast evaluation of ABS placement solutions.}
Our main contributions are summarized as follows:
\begin{itemize}
    \item \textit{Spatial Deep Learning (SDL)\footnote{We reuse the term SDL as in \cite{2019-Yu-Spatial} to refer to the general method of learning spatial characteristics by DL, although different problem setup and DL architecture are considered here.} for Coverage Prediction}: 
    An encoder-decoder type of deep-NN (DNN) with careful incorporation of domain knowledge is proposed.
    First,
    we use three \textit{grid-maps} to quantize and represent the location patterns of ABSs, GUs, and covered GUs (CGUs), respectively, with the first two as input and the last one as output of the DNN.
    These grid-maps endow the DNN with 1) input/output \textit{dimension invariance} with the number of ABSs/GUs; and 2) input/output \textit{permutation invariance} with ABS/GU indexes.
    These invariance properties significantly reduce the learning burden of DNN and 
    also render more flexibility and robustness to scenarios where the number of ABSs/GUs dynamically changes on site.
    Second, 
    tailored design techniques are proposed including \textit{binary mask processing} and \textit{element-wise binary classification}, which effectively boost the training efficiency.
    \item \textit{MAP-Elites as Quality-Diversity Search Engine}: 
    Based on the trained DNN emulator, we further propose the \textit{SDL-based MAP-Elites (SDL-ME)} algorithm.
    First, based on MAP-Elites, the search over the original variable space (of $O(N)$ dimensions) is effectively reduced to that over a low-dimensional (e.g., two in our proposed design) feature space, which flexibly tradeoffs between \textit{complexity reduction} and \textit{solution diversity}, thus encouraging more efficient search for better-quality solutions.
    \rev{Second, the SDL-based emulator captures the site-specific ABS-GU coverage states and enables \textit{virtual exploration} of a much larger part of the search space compared with direct on-site trials and errors, thus leading to potentially better solutions. }
    \item \textit{Top-$k$ Mechanism and Planning-Exploration-Serving (PES) Scheme}: 
    The SDL-ME based planning helps sift out $k$ top performing candidate solutions (in terms of emulator-predicted coverage rate), which are then explored and validated on site to further elect the best performing solution (in terms of actual coverage rate) for ABS placement to serve GUs in the current time period.
    Such a top-$k$ mechanism and PES scheme seamlessly amalgamate the \textit{emulator-based planning} and \textit{on-site exploration/serving}, thus significantly reducing the cost of extensive on-site search.
    Moreover, the measurements for the top-$k$ candidate solutions effectively compensate for the quantization/prediction errors due to model approximation.
    \end{itemize}
    
Numerical results demonstrate that the proposed approach significantly outperforms the benchmark DRL-based method and other two baselines in terms of average coverage rate, training time and/or sample efficiency.
Moreover, with one-time training, our proposed method can be applied in \textit{dynamic scenarios} where the number of ABSs/GUs changes on site and/or with different/varying GU speeds, which is more robust and flexible compared with conventional DRL methods. 

\section{System Model}\label{sec:system_model}

\begin{figure}[!t]
\centering
\includegraphics[width=1\linewidth,  trim=0 0 0 0,clip]{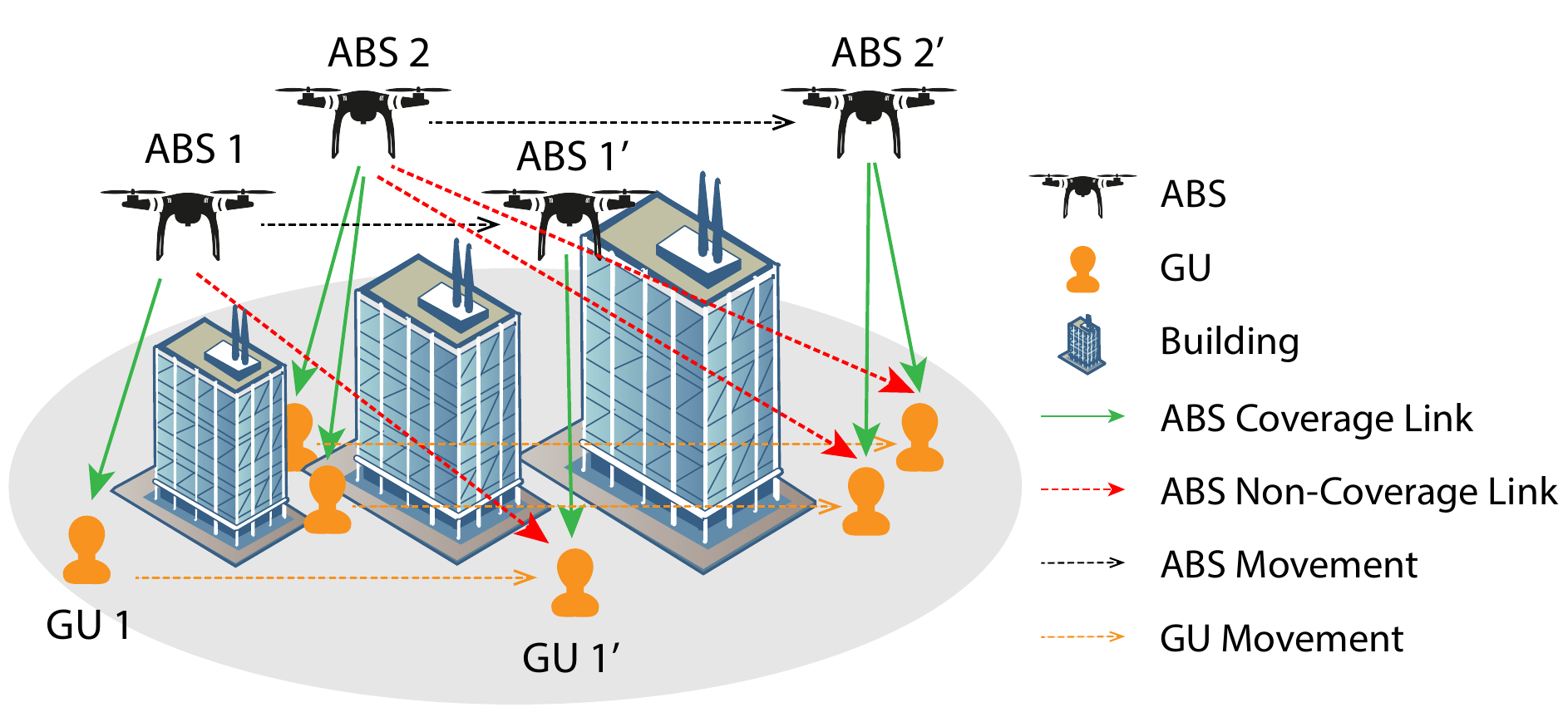}
\caption{\rev{ABS movement optimization to cover mobile GUs in an environment with site-specific blockages (e.g., GU 1 previously covered by ABS 1 may move and become uncovered due to blockage, whereby ABS 1 may move adaptively to cover it again).}\vspace{-2ex}}
\label{fig:3D_illustration}
\end{figure}

Consider a UAV-aided communication system with $N$ UAV-mounted ABSs to serve a group of $M$ mobile GUs in a $D\times D$ m$^2$ square area with site-specific blockages, as illustrated in Fig. \ref{fig:3D_illustration}. 
For the purpose of exposition, the blockages are exemplified using a collection of $L$ building blocks (BBs), each with a $D_{w}\times D_{w}$ m$^2$ square projection shape and a random height $h_{w}[l]$, $l\in\mathcal{L}\triangleq\{1, \dots, L\}$. 
In this work, we focus on the access network where ABSs strive to provide data communication coverage for GUs, and assume for simplicity that there exists a backhaul network among ABSs.\footnote{The ABS-ABS channel is more likely to be LoS-dominated and hence suitable for establishing a connected backhaul network.}
\rev{Table \ref{TableSingle} summarizes the acronyms for easier reference.}

\subsection{ABS/GU Mobility Model}\label{SectionMobility}
Consider a typical ABS movement \textit{trial} of duration $\Delta T$ s, which is discretized into $I$ equal-length time \textit{steps}, each of duration $\Delta \tau=\Delta T/I$. 
For simplicity, assume that ABSs fly at a fixed altitude\footnote{
\rev{A statistically optimal ABS altitude can be found based on the channel statistics, in order to maximize its average ground coverage range \cite{2014-Hourani-Optimal}. Real-time three-dimentional (3D) ABS movement is left for future work.}
}
of $h_p$ m and GUs move on the ground with a hand-held height of $h_q$ m.
Moreover, assume that the GUs move at a constant speed $V_q$ m/s but with a random direction in each step.\footnote{Our proposed method is general and can be similarly applied to other GU mobility models.}
Assume that the locations of GUs and ABSs in each step are known and reported via separate control links to a central planning agent, which could be mounted on one of the UAVs or a ground vehicle station.
Denote $\boldsymbol{p}^{(i)}[n]=(x^{(i)}_{p}[n], y^{(i)}_{p}[n])$ as the horizontal position of ABS $n\in\mathcal{N}\triangleq\{1, \dots, N\}$ at step $i\in \mathcal{I}\triangleq\{1, \dots, I\}$.
Similarly, denote $\boldsymbol{q}^{(i)}[m]=(x^{(i)}_{q}[m], y^{(i)}_{q}[m])$ as the horizontal position of GU $m\in\mathcal{M}\triangleq\{1, \dots, M\}$ at step $i\in\mathcal{I}$.
Denote $\mathcal{P}^{(i)}\triangleq\{\boldsymbol{p}^{(i)}[n]|n\in\mathcal{N}\}$ or $\mathcal{Q}^{(i)}\triangleq\{\boldsymbol{q}^{(i)}[m]|m\in\mathcal{M}\}$ as the location set at step $i\in\mathcal{I}$ for ABSs or GUs, respectively.\footnote{With slight abuse of notations, we might omit the superscript $(i)$ when referring to a general location without emphasis on the time step $i$. For example, $\boldsymbol{p}[n]$ indicates the position of ABS $n$ in general.}

\begin{table}[t]\small
	\renewcommand\arraystretch{0.65} 
	\centering
	\caption{\rev{List of frequently used acronyms.}}
	\label{TableSingle}
	\begin{tabular}{c p{0.32\linewidth}|c p{0.36\linewidth}}
		\toprule
		UAV & unmanned aerial vehicle & ABS & aerial base station \\
		\midrule
		GU & ground user & CGU & covered ground user  \\
		\midrule
		SDL & spatial deep learning & ME & MAP Elites \\
				\midrule
		NN & neural network & RL & reinforcement learning \\
				 				\midrule
		 DNN & deep NN & CNN & convolutional NN \\
		 		 				\midrule
		 DRL & deep RL &TD3 & twin delayed deep deterministic policy gradient \\
		 				\midrule
		 LoS & line-of-sight &NLoS & none line-of-sight \\
		 		 				\midrule
		 EA & evolutionary algorithm & PES & planning-exploration-serving \\
		 		 		 				\midrule
		 BB & building block & SNR & signal-to-noise ratio \\
		 		 		 		 				\midrule
		 CR & coverage rate & NM &naive mutation \\
		 	\midrule
		 ACR & average coverage rate & GES &grid exhaustive search\\
		\bottomrule
	\end{tabular}
\end{table}

Assume that each ABS can adjust its own moving speed as needed, subject to a maximum speed constraint of $V^{\max}_p$ m/s. Denote $\Vert \cdot \Vert$ as the Euclidean norm. Then the ABS positions in consecutive time steps are constrained by the maximum moving distance, i.e., 
\begin{align}
    \lVert \boldsymbol{p}^{(i)}[n] - \boldsymbol{p}^{(i-1)}[n] \rVert \le V^{\max}_p \cdot \Delta \tau, \forall i\in\mathcal{I}, n\in\mathcal{N},\label{eq:moving_speed}
\end{align}
Furthermore, to avoid collisions between ABSs, a minimum security distance $d_{\min}$ is imposed,\footnote{\rev{Due to discrete time consideration, collision may still happen within a particular time step if two UAVs' trajectories happen to cross over. One practical remedy is to equip the UAVs with collision-avoiding sensors and flight control units to locally detect and bypass each other.}} i.e.,
\begin{equation}\label{eq:min_distance}
    \lVert \boldsymbol{p}^{(i)}[a] - \boldsymbol{p}^{(i)}[b] \rVert \ge d_{\min}, ~\forall~i\in\mathcal{I},~a,b\in\mathcal{N}, a\neq b.
\end{equation}
Finally, we focus on the outdoor scenario within bounded area in this paper.
Denote $\mathcal{A}$ as the considered area in the horizontal plane at alititude $h_p$, and $\mathcal{B}\subset \mathcal{A}$ as the region occupied by obstacles. The following constraint is thus imposed, i.e.,
\begin{align}
    \boldsymbol{p}^{(i)}[n]\in\mathcal{A}\setminus \mathcal{B}, \forall i\in\mathcal{I}, n\in\mathcal{N}.\label{eq:area}
\end{align}

\subsection{Site-Specific LoS/NLoS Channel Model}

Consider downlink communication from ABSs to GUs, while the proposed approach can be similarly applied to uplink communication. Assume that each ABS or GU is equipped with omni-directional antenna with unit gain.
\rev{For simplicity and the purpose of exposition, assume that a total spectrum bandwidth of $B$ Hz is equally divided into $N$ orthogonal bands each with $B/N$ Hz and allocated to one ABS. Since each ABS might have limited resource/capability, we assume that each ABS can support a maximum number of $M_{\textrm{max}}$ GUs, and consider a simple GU-ABS association rule based on the constrained K-means clustering algorithm \cite{ConstrainedKmeans} with maximum size $M_{\textrm{max}}$ for each cluster.\footnote{\rev{The GU-to-ABS association can be formulated as the cluster assignment sub-problem in \cite{ConstrainedKmeans}, which is equivalent to a Minimum Cost Flow linear network optimization problem that can be solved efficiently (https://pypi.org/project/k-means-constrained/).}} Assume that the transmit power $P_\text{t}$ Watt and bandwidth at each ABS $n$ is equally allocated among its associated $M_n$ GUs, respectively. As a result, the instantaneous signal-to-noise ratio (SNR) received by GU $m$ from its associated ABS $n$ is given by
\begin{equation}\label{gamma}
\gamma_{m,n}\triangleq \frac{g_{m,n} P_\text{t}/M_n}{N_0 (B/N)/M_n} =\frac{g_{m,n} N P_\text{t}}{N_0 B}=g_{m,n} N \gamma_\text{t},
\end{equation}
where the receiver noise is assumed to be additive white Gaussian noise (AWGN) with power spectrum density $N_0$ (W/Hz), $\gamma_\text{t}\triangleq P_\text{t}/(N_0 B)$ denotes the nominal transmit SNR with a single ABS,} and $g_{m,n}\triangleq \bar g_{m,n} \xi_{m,n}$ denotes the instantaneous channel power gain, with $\bar g_{m,n}$ denoting the average channel power and $\xi_{m,n}$ accounting for small scale fading with unit average power.

Due to site-specific blockages, the ABS-GU channel could be in either LoS or NLoS propagation condition depending on whether there are obstacles in between.
Therefore, the average channel power gain between GU $m$ and ABS $n$ can be expressed as
\begin{align}\label{probLOS}
\bar g_{m,n}\triangleq
\begin{cases}
\bar g_{\textrm{L}}(\boldsymbol{p}[n], \boldsymbol{q}[m]), & \quad \textrm{no obstacles in between;}\\
\bar g_{\textrm{NL}}(\boldsymbol{p}[n], \boldsymbol{q}[m]), & \quad \textrm{otherwise,}
\end{cases}
\end{align}%
where $\bar g_{\textrm{L}}$ and $\bar g_{\textrm{NL}}$ denote the average channel power gains of the LoS and NLoS channels, respectively.\footnote{As a preliminary study, we adopt the urban macro formulas in 3GPP \cite{3GPP} as the underlying path-loss model in our simulations.}
\rev{
Regarding small-scale fading, for the LoS case, consider the angle-dependent Rician fading channel with factor $K_{m,n}$ given by \cite{TWCChangshengYou3DRicianFadingUAV}
\begin{equation}\label{RicianK}
    K_{m,n} = A_{1}\textrm{exp}(A_{2}\theta_{m,n}),
\end{equation}
where $A_{1}$ and $A_{2}$ are constant coefficients, and $\theta_{m,n}\triangleq \arctan\frac{h_p-h_q}{\lVert \boldsymbol{p}[n] - \boldsymbol{q}[m] \rVert}$ is the elevation angle of ABS $n$ as seen by GU $m$.
Then we have $K_{\textrm{min}} \leq K \leq K_{\textrm{max}}$, where $K_{\textrm{min}} = A_{1}$ and $K_{\textrm{max}} = A_{1}e^{A_{2}\pi/2}$. On the other hand, for the NLoS case, Rayleigh fading is considered which is a special case of Rician fading with $K_{m,n}=0$.\footnote{
\rev{Note that the channel model in \eqref{probLOS} and \eqref{RicianK} only serves as the underlying ground truth model used in the simulation studies. Our proposed scheme does not assume knowledge of the geometric environment map or instantaneous channel state information, but instead relies on on-site connectivity/throughput measurements, similar to the procedure in \cite{YangUAV}.
}}

Due to small scale fading, the instantaneously received SNR might fall below a certain required level $\bar\gamma$ and cause communication outage, with the outage probability $P_{\text{out},m}\triangleq \textrm{Pr}\{\gamma_{m,n}<\bar\gamma\}$. As a result, the average throughput of GU $m$ can be given by
\begin{equation}\label{Rate}
    R_m \triangleq \frac{(1-P_{\text{out},m} ) B}{NM_n}\log_2 (1+\bar\gamma).
\end{equation}
It can be seen from \eqref{Rate} and \eqref{gamma} that the GU’s throughput in bits/second (bps) depends on the system bandwidth $B$, the transmit SNR $\gamma_\text{t}$, the number of ABSs $N$, the number of associated GUs $M_n$ at each ABS $n$, and the channel power $g_{m,n}$. In particular, when many GUs are associated with one ABS, each of these GUs would have a small throughput due to limited resources per ABS. This would thus encourage other ABSs to come over and help serve these GUs. 
Denote $\bar R$ as the threshold of required throughput level.
We can then define a \textit{coverage indicator} for GU $m$ in step $i$ as
\begin{equation}\label{Cm}
C_m^{(i)}\triangleq\begin{cases}
1, & \textrm{if} \ R_m^{(i)}\geq \bar R, \\
0, & \textrm{otherwise.}%
\end{cases}%
\end{equation}}%
Finally, the \textit{coverage rate (CR)} of all GUs in step $i$ is given by
\begin{equation}\label{CR}
    \lambda^{(i)}\triangleq \frac{1}{M}\sum\nolimits_{m\in\mathcal{M}} C_m^{(i)}.
\end{equation}

\section{Problem Formulation}\label{sec:problem_formulation}
We aim to maximize the average coverage rate (ACR) $\bar\lambda$ over the entire trial by muti-ABS movement optimization, as given by
\begin{equation*}
    \begin{aligned}
        \text{(P1):}~\max_{\mathcal{P}^{(i)}, i\in\mathcal{I}}~&\bar\lambda\triangleq\frac{1}{I}\sum\nolimits_{i \in\mathcal{I}}\lambda^{(i)}\\
        \text{s.t.} ~& \eqref{eq:moving_speed}, \eqref{eq:min_distance}, \eqref{eq:area}, \eqref{Cm}~\text{and}~\eqref{CR}.
    \end{aligned}
\end{equation*}%
\rev{The main difficulties to solve problem (P1) include: (1) multi-ABS placement for maximizing multi-GU coverage, (2) site-specific LoS/NLoS channel, (3) GU mobility, and (4) dynamic network conditions. The correlation of these difficulties lies in the following aspects: 1) multi-ABS placement is NP-hard with the number of ABSs/GUs for obtaining the optimal solution; 2) the problem is further complicated by site-specific channel states of all ABS-GU links,\footnote{\rev{For $N$ ABSs and $M$ GUs, there are $MN$ links each with two possible states of LoS/NLoS condition due to site-specific blockages, resulting in an irregular and enormous state space with $2^{MN}$ combinations.}}
which make it difficult to analytically model/predict/optimize the overall coverage performance; 3) GU mobility requires timely ABS movement solutions that should be found within limited time and with limited chances of on-site measurements, thus placing more restrictions to the above problem; and 4) the solution is desired to be robust and flexible under dynamic network conditions where the number of ABSs/GUs changes on site and/or with different/varying GU speeds.}

\section{Site-Specific Movement Optimization of Aerial Base Stations}\label{sec:proposed_algorithm}

\rev{To circumvent the above difficulties, we attempt a different approach other than DRL, and propose the novel SDL-ME method to solve the site-specific multi-ABS movement optimization problem.
Specifically, a three-level time hierarchy is first proposed in Section \ref{SectionHierachy}, which partitions problem (P1) into a time series of ABS placement sub-problems, each with given GU locations.
For each sub-problem, the goal is to find the optimal ABS location set to maximize the coverage rate for the current GU location set. 
Nevertheless, due to the enormous and irregular state and/or action space as well as the NP-hard nature of the problem, conventional search/optimization algorithms \cite{2011-Pham-Intelligent} might incur prolonged computational/time overhead, require excessive one-site measurement/cost, and/or fail to find desirable solutions.
To this end, we first propose a \textit{SDL-based emulator} in Section \ref{SectionEmulator}, which can predict the coverage states of all GUs in the actual environment and assist fast evaluation of ABS placement solutions. The emulator is based on an encoder-decoder type of DNN, which is designed with careful incorporation of domain knowledge to reduce learning complexity while rendering 
more flexibility and robustness to scenarios with dynamic network changes.
Furthermore, in Section \ref{SectionMAP}, we employ one of the state-of-the-art search algorithms named MAP-Elites \cite{2015-MAP-Elites} as the \textit{quality-diversity search} engine, which promotes both search efficiency and solution diversity for further achieving global optimality.
Finally, a novel \textit{PES scheme} is proposed in Section \ref{PES} to seamlessly amalgamate the emulator-based planning and on-site exploration/serving, which complement each other in reducing both the performance loss due to model approximation, and the need for extensive on-site trials and errors.
}

\subsection{Three-Level Time Hierarchy}\label{SectionHierachy}
\rev{A \textit{trial-period-step} time hierarchy is proposed to partition problem (P1) into a time series of ABS placement sub-problems.
Ideally, each time step requires a dedicated ABS placement sub-problem, in order to make the most timely adaptation to GU location updates.
However, such granularity could be computationally prohibitive, and sometimes unnecessary in cases with low/moderate GU mobility.
To this end, we introduce an intermediate time level named \textit{period}, each lasting for several steps and corresponding to one ABS placement sub-problem and thus one cycle of planning, exploration and serving.
The duration of one period can be flexibly chosen in order to trade off between computational overhead, quality of solutions, and timeliness for GU coverage.
}

\begin{figure}[!t]
\centering
\includegraphics[width=0.95\linewidth,  trim=30 50 10 20,clip]{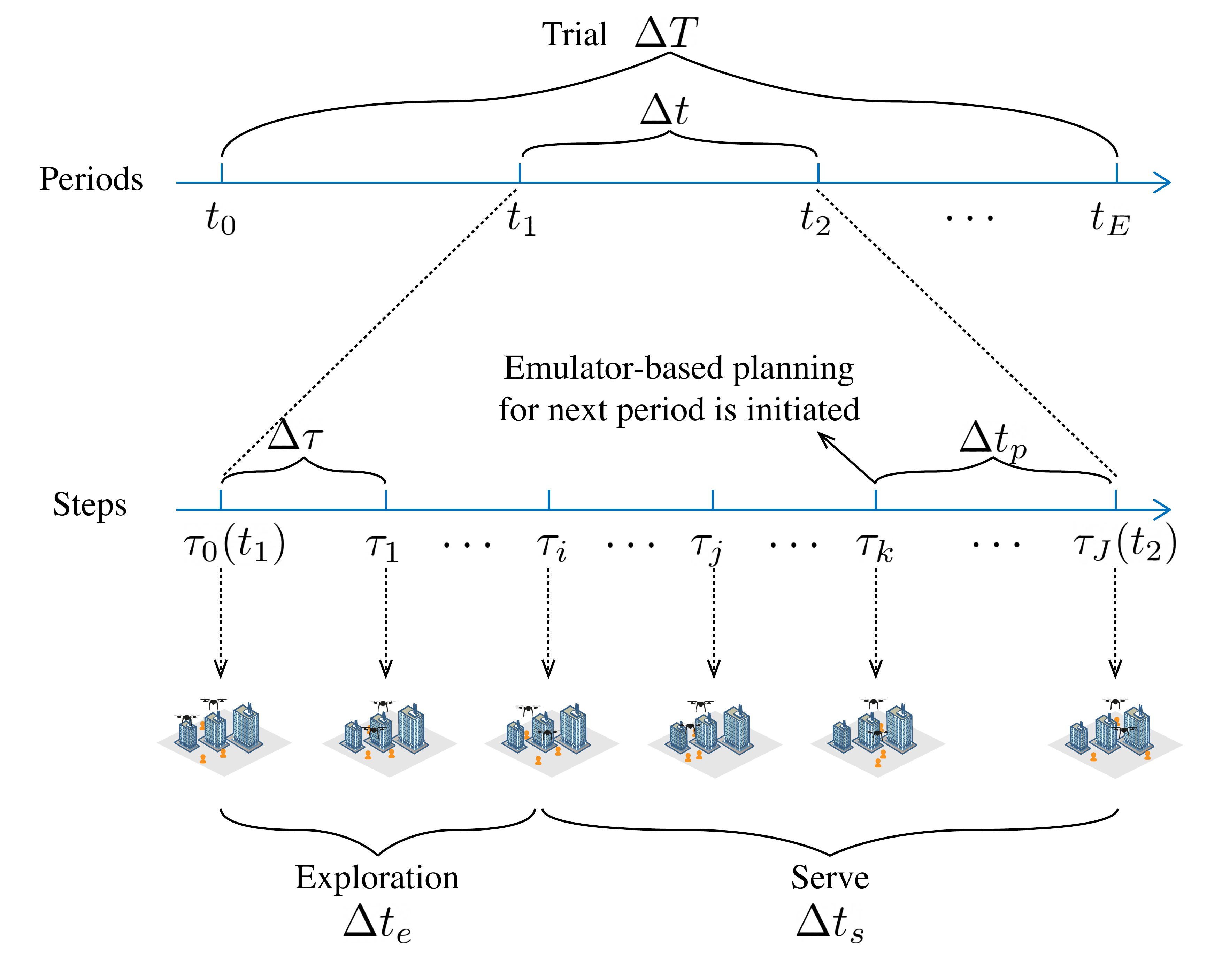}
\caption{Three-level \textit{trial-period-step} time hierarchy for ABS movement optimization.\vspace{-2ex}}
\label{fig:hierarchy}
\end{figure}

Specifically, one trial for problem (P1) is equally divided into $E$ \textit{periods}, where each period has $J\triangleq I/E$ steps and thus lasts for $\Delta t\triangleq\Delta T/E= J\cdot\Delta \tau$ s, as illustrated in Fig. \ref{fig:hierarchy}.
As a preliminary study, we consider the scenario with low/moderate GU mobility, and assume that the GU distribution between two consecutive periods will not be drastically different.
Each period consists of two non-overlapping phases, i.e., exploration and serving, each with a duration of $\Delta t_{e}$ s and $\Delta t_{s}$ s, respectively, with $\Delta t_{e}+\Delta t_{s}=\Delta t$. 
For a target period (e.g., $t_1\sim t_2$ in Fig. \ref{fig:hierarchy}), there is a prerequisite planning procedure of duration $\Delta t_p$ s ($\Delta t_p\leq \Delta t$) beforehand, which takes the newly reported GU location information (e.g., at time instant $t_1-\Delta t_p$) as input, and outputs the candidate ABS placement solutions to feed to the target period.
Next, the target period tries out these candidate solutions by on-site measurements in the exploration phase, and finds the best solution to be used in the serving phase. More detailed design of exploration and serving will be discussed later in Section \ref{PES}. 
Next, we focus on the proposed emulator design and the emulator-based planning procedure first.

\subsection{SDL-based Emulator Design}\label{SectionEmulator}
\rev{The overall architecture of the proposed DNN-based emulator is shown in Fig. \ref{fig:emulator}, whose main novelty lies in the following joint design considerations: 1) grid-map based quantization of ABS/GU locations, 2) binary mask processing and the formulation of element-wise binary classification task, 3) and choosing a suitable DNN model with state-of-the-art performance to fit and solve our considered problem. More details are provided in the sequel.
}

\subsubsection{Grid-Map based Quantization of ABS/GU Locations}
\rev{First of all, the DNN input/output structure needs careful design.}
The emulator needs to predict the coverage status of all GUs efficiently, given the ABS/GU locations in a site-specific area.
To achieve this goal, a common choice of input could be the coordinates of all ABS and GU locations, while the output could be the coverage rate directly. 
\rev{However, the dimension and associated complexity of such input representation increase with $N$ and $M$. In addition, it typically cannot accommodate changing number of ABS/GUs since the input dimension of DNN is typically pre-determined before training. Moreover, capturing spatial correlations with scalar-type location variables is difficult even for DNN models, and the implicit ordering among the coordinates of ABSs/GUs adds unnecessary learning burden to the DNN. In the meanwhile, predicting the coverage rate directly might abandon rich latent knowledge about the spatial distribution of covered GUs and thus lead to poor performance.}

In this paper, we propose to employ three \textit{grid-maps} to quantize and represent the location patterns of ABSs, GUs, and covered GUs (CGUs), respectively, with the first two as input and the last one as output of the emulator.
Specifically, we partition the studied horizontal area into $K$-by-$K$ grid regions, denoted by $G_{i,j}$, $i,j\in\mathcal{K}\triangleq\{1,\dots,K\}$.
Note that the choice of $K$ needs to balance between the pattern resolution and model complexity.\footnote{For simplicity, the same resolution $K$ is chosen for both ABS and GU patterns, which in general can be set differently as needed.}
However, it is found that a moderately large $K$ would suffice the model accuracy requirement for our experiments.
As a result, we can define the \textit{ABS pattern} as a $K$-by-$K$ matrix $P_p\triangleq [f^p_{i,j}]_{K\times K}\in\mathbb{Z}^{K\times K}$, whose $(i,j)$-th element is given by
\begin{equation}\label{eq:GU_pattern}
    f^p_{i,j}\triangleq\sum\nolimits_{n\in\mathcal{N}} \mathbbm{1} \{\boldsymbol{p}[n]\in G_{i,j}\},
\end{equation}
which represents the number of ABSs in that grid, with $\mathbb{Z}$ denoting the set of non-negative integers. The indicator function $\mathbbm{1}\{S\}$ equals to 1 when the statement $S$ is true, and 0 otherwise.
Similarly, we can define the \textit{GU pattern} as $P_q\triangleq [f^q_{i,j}]_{K\times K}\in\mathbb{Z}^{K\times K}$ and \textit{CGU pattern} as $P_{\hat{q}}\triangleq [f^{\hat{q}}_{i,j}]_{K\times K}\in\mathbb{Z}^{K\times K}$, whose $(i,j)$-th element is given by
\begin{equation}
    f^q_{i,j}\triangleq\sum\nolimits_{m\in\mathcal{M}} \mathbbm{1} \{\boldsymbol{q}[m]\in G_{i,j}\},
\end{equation}
\begin{equation}\label{eq:CGU_pattern}
    f^{\hat{q}}_{i,j}\triangleq\sum_{\boldsymbol{q}[m]\in G_{i,j}, m\in\mathcal{M}} C_m,
\end{equation}
respectively, where $C_m$ is the coverage indicator for GU $m$ defined in \eqref{Cm}. 
\rev{A toy example is shown in the upper part of Fig. \ref{fig:emulator} to demonstrate how to obtain the quantized CGU pattern $P_{\hat{q}}$.}

\begin{remark}
\rev{Based on such grid-map design, the emulator input/output dimensions become \textit{invariant} to $N$ and $M$ but instead determined by the granularity $K$, which renders more flexibility and robustness to scenarios where the number of ABSs/GUs dynamically changes on site.
Moreover, the ordering of ABS or GU indexes becomes \textit{permutation invariant}, which also significantly reduces the learning burden of the underlying DNN.
Last but not least, such grid-map based input/output design enables us to reuse the state-of-the-art Convolutional NN (CNN) with image encoder-decoder structure, e.g., Attention-UNet \cite{2018-Oktay-AttentionUNet}, in order to fit and solve our considered problem.}
\end{remark}

\subsubsection{Binary Mask Processing}
\begin{figure*}[!t]
\centering
\includegraphics[width=0.9\textwidth]{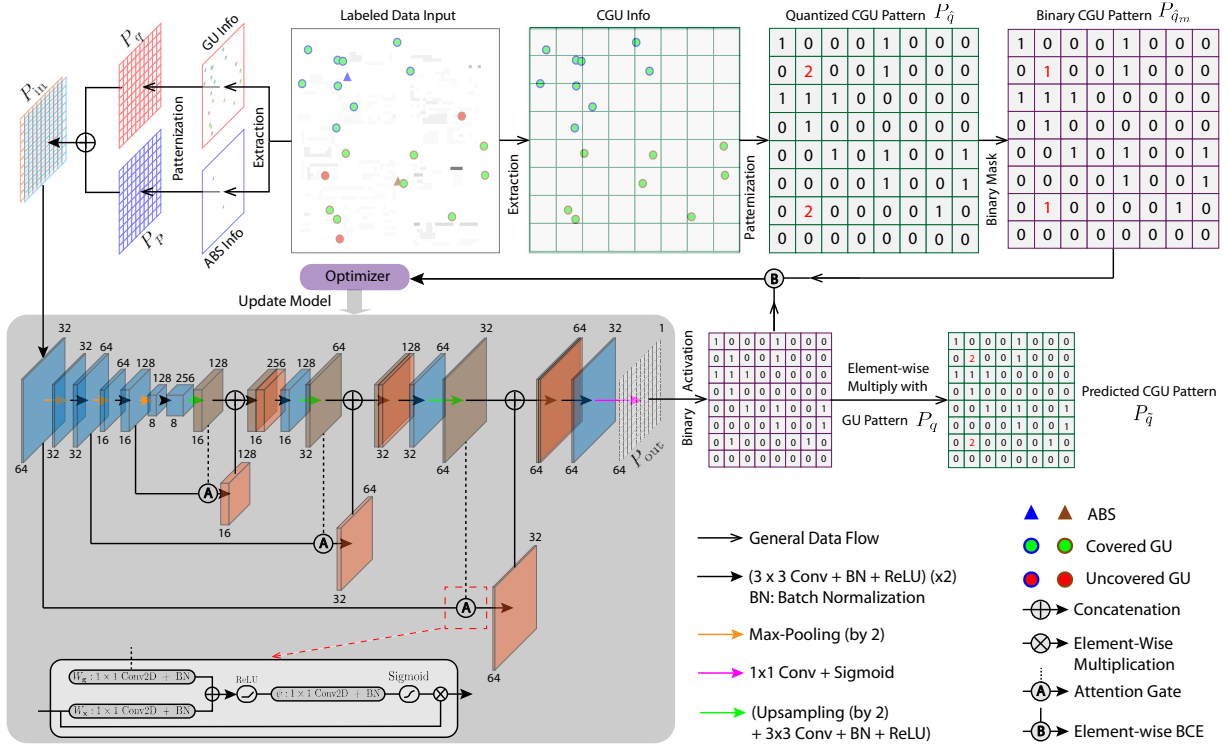}
\caption{\rev{The overall architecture and training workflow of the proposed environment emulator.}\vspace{-2ex}}
\label{fig:emulator}
\end{figure*}

\rev{In the above grid-map representation, each element in the pattern matrix ($P_p$, $P_q$ or $P_{\hat{q}}$) represents the number of entities in that grid, which takes integer values from 0 to $N$ or $M$.
By taking $P_p$ and $P_q$ as input and $P_{\hat{q}}$ as output, the learning task of the emulator model becomes a \textit{multi-dimensional (MD) regression} problem, which is still challenging for large values of $N$ and/or $M$.
To this end, we propose to apply a \textit{binary mask processing} operation on each element of the output pattern $P_{\hat{q}}$ to get the masked CGU pattern $P_{\hat{q}_m}$, i.e., treating each grid as a whole and predicting that GUs in each grid either all being covered or not.
Specifically, the $(i,j)$-th element of $P_{\hat{q}_m}$ is given by $f^{\hat{q}_m}_{i,j}\triangleq \mathbbm{1} \{f^{\hat{q}}_{i,j} > 0\}$.\footnote{\rev{For the toy example in Fig. \ref{fig:emulator}, two CGUs reside in the (2, 2)-th grid, and hence we have $[P_{\hat{q}}]_{2,2}=2$ and $[P_{\hat{q}_m}]_{2,2}=1$.}}
Although such a binary operation coarsens the emulator output, it further reduces the learning burden of the underlying DNN by transforming the original MD regression task into an easier \textit{MD binary classification} task.}
As a result, the concatenation of the ABS pattern $P_p$ and GU pattern $P_q$, denoted as $P_{\text{in}}$, is taken as the \textit{input} of the DNN model $\Omega$, while $P_{\hat{q}_m}$ is treated as the \textit{output label}.

After training, the DNN model $\Omega$ produces a $K$-by-$K$ matrix as the \textit{output}, given by 
\begin{equation}
    P_{\text{out}} \triangleq\Omega(P_{\text{in}};\theta) \in\mathbb{P}^{K\times K},
\end{equation}
where $\mathbb{P}\triangleq \{x|0\leq x\leq 1\}\in \mathbb{R}$ denotes the real-valued set ranging from 0 to 1, and $\theta$ represents the model parameters.
Note that the direct output of the DNN model $\Omega$ takes a probability measure. Specifically, the $(i,j)$-th element of $P_{\text{out}}$, denoted by $f_{i,j}^{\text{out}}$, represents the \textit{predicted probability} that GUs in grid $(i,j)$ are all covered.
Based on such probability and the original GU pattern $P_q$, the emulator outputs the \textit{final predicted CGU pattern} $P_{\tilde{q}}\in\mathbb{Z}^{K\times K}$, whose $(i,j)$-th element is given by
\begin{equation}
f^{\tilde{q}}_{i,j}\triangleq f^{q}_{i,j}\cdot \mathbbm{1} \{f^{\text{out}}_{i,j} > \eta\},
\end{equation}
\rev{where $\eta=0.5$ is a probability threshold used in the output layer\footnote{\rev{The Sigmoid function is used to output either 0 or 1, which is  an “S”-shaped curve that is rotationally symmetric about the mid-value of 0.5.}} to determine the coverage status of a grid.}

\subsubsection{Binary Cross Entropy for Element-wise Binary Classification}
Finally, we introduce the metric used for training the DNN model $\Omega$, i.e., the Binary Cross Entropy (BCE) \cite{2016-Goodfellow-DL}. Based on the above discussions, it is desired that the output $P_{\text{out}}$ of $\Omega$ matches the output label $P_{\hat{q}_m}$ as closely as possible, which amounts to a $K$-by-$K$ dimensional binary classification task.
Normally, BCE is employed as the loss function for binary classification tasks. 
For our considered MD binary classification problem, we define its loss function as the sum of element-wise BCE (E-BCE) between $P_{\text{out}}$ and $P_{\hat{q}_m}$, i.e., 

\begin{small}
\begin{align}\label{eq:loss_function}
    &L_{\textrm{BCE}}(\theta) \triangleq\notag\\
    &- \mathbb{E}\bigg[\frac{1}{K^2}\sum^K_{i=1}\sum^K_{j=1} \bigg ( f^{\hat{q}_m}_{i,j}\cdot\log f^{\text{out}}_{i,j} + (1-f^{\hat{q}_m}_{i,j})\cdot\log(1-f^{\text{out}}_{i,j})\bigg)\bigg].%
\end{align}%
\end{small}%
The training for the DNN model $\Omega$ then tries to minimize the above loss function using typical optimization methods, by feedforward inference to get a prediction $P_{\text{out}}$ and back-propagation of loss values to update the parameter $\theta$.

\begin{remark}
    \rev{The grid-map quantization and binary mask processing, along with the E-BCE metric, effectively turn the original coverage rate prediction task into an \textit{element-wise binary classification} task, which greatly alleviates the difficulty for model approximation.}
\end{remark}


\subsubsection{Capturing Coverage Correlations with Attention-UNet}
In order to choose a suitable DNN architecture, we first analyze the \textit{inherent properties} of the ABS, GU and CGU patterns.
1) All three patterns take an image shape, which enable us to reuse the state-of-the-art CNN with image encoder-decoder structure. 
2) The CGU set is always a subset of the GU set, which indicates a strong connection between the CGU pattern $P_{\hat{q}}$ and GU pattern $P_{q}$.
3) The coverage status of a GU is determined by its surrounding environment and the relative locations of nearby ABSs, and therefore an ABS in nearby grids will increase the chance of the GU to be covered.

\rev{Based on the above analysis, a specialized CNN architecture is desired to cater for the design considerations and task formulation discussed earlier. Note that the typical use of CNN is on classification tasks, where the output to an image is a single-class label. However, in our considered DNN model, we need to determine the coverage (activation) status of a GU in a certain grid (pixel). As a result, the desired DNN output should include localization, i.e., a class label is supposed to be assigned to each pixel. Therefore, our emulator learning strongly resembles the image segmentation task,\footnote{Image segmentation is the process of partitioning a digital image into multiple image segments, which is typically used to locate objects and boundaries (lines, curves, etc.) in images.} which is a MD classification problem and considerably more difficult than image classification. To address the above challenge, we pick Attention-UNet \cite{2018-Oktay-AttentionUNet} as the basis of our underlying NN model, which is a special type of CNN with delicate designs to cater for the image segmentation task, including the UNet structure, skip connections, and attention gate. }
An exemplary implementation of the Attention-UNet structure is shown in the lower part of Fig. \ref{fig:emulator}. 
The \textit{Attention-UNet} model first uses convolutional layers to progressively extract higher dimensional pattern representations by processing local information layer by layer, which eventually separates elements of the given pattern in a high dimensional space according to their activation statuses.
Through this \textit{sequential downsampling encoding}, the model can extract pattern features at multiple resolution scales, with coarse feature maps extracted at lower layers to capture contextual information and highlight the activation status and location of target regions.
Then, through a reverse process of \textit{upsampling decoding}, the feature maps are combined comprehensively to predict/reconstruct desired output patterns.
Finally, \textit{skip connections} are applied in the above encoding-decoding process (UNet structure) to directly connect coarse feature maps with fine-level dense predictions to yield better results.
In particular, \textit{attention gates (AGs)} are integrated in this model to automatically focus on target structures while suppressing the feature responses in irrelevant regions with a low computational cost, as shown in Fig. \ref{fig:emulator}. 


\subsubsection{Data Collection and Training}\label{DataCollection}
In order to obtain a robust DNN model, it is generally required to collect a sufficient number of samples or labeled data points, each with given $P_p$, $P_q$ and $P_{\hat{q}}$.
For such purpose, the ABSs move coordinately towards their planned destinations in each period while following the movement constraints \eqref{eq:moving_speed} - \eqref{eq:area}. As each ABS moves towards its destination(s), $J$-step connectivity/throughput measurements are performed along the path for each ABS-GU link, which together yield the CGU pattern $P_{\hat{q}}$ per step.
For the initial dataset collection, the planned ABS positions in each period are obtained by either the constrained random (cRandom) or constrained K-means (cKMeans) placement strategy.
The cRandom strategy randomly samples new ABS positions while satisfying the constraints \eqref{eq:moving_speed} - \eqref{eq:area}.
On the other hand, the cKMeans strategy first employs the K-means centering algorithm with random seed according to GU locations, and then check and retain the resulted ABS placement solution if the constraints \eqref{eq:moving_speed} - \eqref{eq:area} are satisfied. 


Once the data collection is done, we partitioned the collected samples into a training set and a validation set. 
We use the training set to train the model for minimizing the loss function (\ref{eq:loss_function}) with typical optimization methods, e.g., stochastic gradient decent (SGD), and use the validation set to periodically validate model accuracy and avoid overfitting.
\rev{Note that the adopted Attention-UNet architecture above is a special type of CNN, whose computational complexity is mainly dominated by all the convolution layers inside the network \cite{CNNtime}, which is given by $O(\Sigma_{l=1}^d n_{l-1} s_l^2 n_l m_l^2)$. Here $l$ is the index of a convolutional layer, $d$ is the depth (number of convolutional layers), $n_l$ is the number of filters (also known as “width”) in the $l$-th layer, $n_{l-1}$ is also known as the number of input channels of the $l$-th layer, $s_l$ is the spatial size (length) of the filter, and $m_l$ is the spatial size of the output feature map. Note that this time complexity applies to both training and testing time, though with a different scale. The training time per sample is roughly three times of the testing time per sample (one for forward propagation and two for backward propagation) \cite{CNNtime}. Regarding the training process, we have adopted the classic SGD method, whose convergence to at least locally optimal point is guaranteed \cite{SGDoxford}, and could be further accelerated by its advanced variant such as Adam \cite{TomLuoAdam}.}

\begin{remark}
    Our proposed emulator is inherently able to make coverage predictions under a \textit{variable number of ABSs/GUs}, thanks to our grid-map based input/output structure and the generalization capability of Attention-UNet.
\end{remark}

\subsection{Emulator-Based Planning}\label{SectionMAP}
Next, we introduce the emulator-based planning to find a set of sub-optimal ABS placement patterns for a specific GU pattern in each period.
The planning procedure can employ any search/optimization algorithm that finds desired ABS placement solutions under given budget of time and computational resource.
For the purpose of exposition, we consider the category of evolutionary algorithms (EAs) \cite{2011-Pham-Intelligent} here and leave other methods for future investigation. 
In the EA parlance, a solution is an \textit{individual} described by a \textit{genotype}. The performance or quality of a solution is called its \textit{fitness}, and the equation or simulation that returns that fitness value is the \textit{fitness function}.
In addition, the process of stochastically producing new genotypes from an existing one is called \textit{mutation}.
For our considered setup with given GU locations in each period, the ABS placement solution resembles an individual with a certain genotype, and its associated CR resembles its fitness. 

\subsubsection{Mutation Operation for ABS Patterns}
\begin{figure}[!t]
\centering
\includegraphics[width=0.95\linewidth]{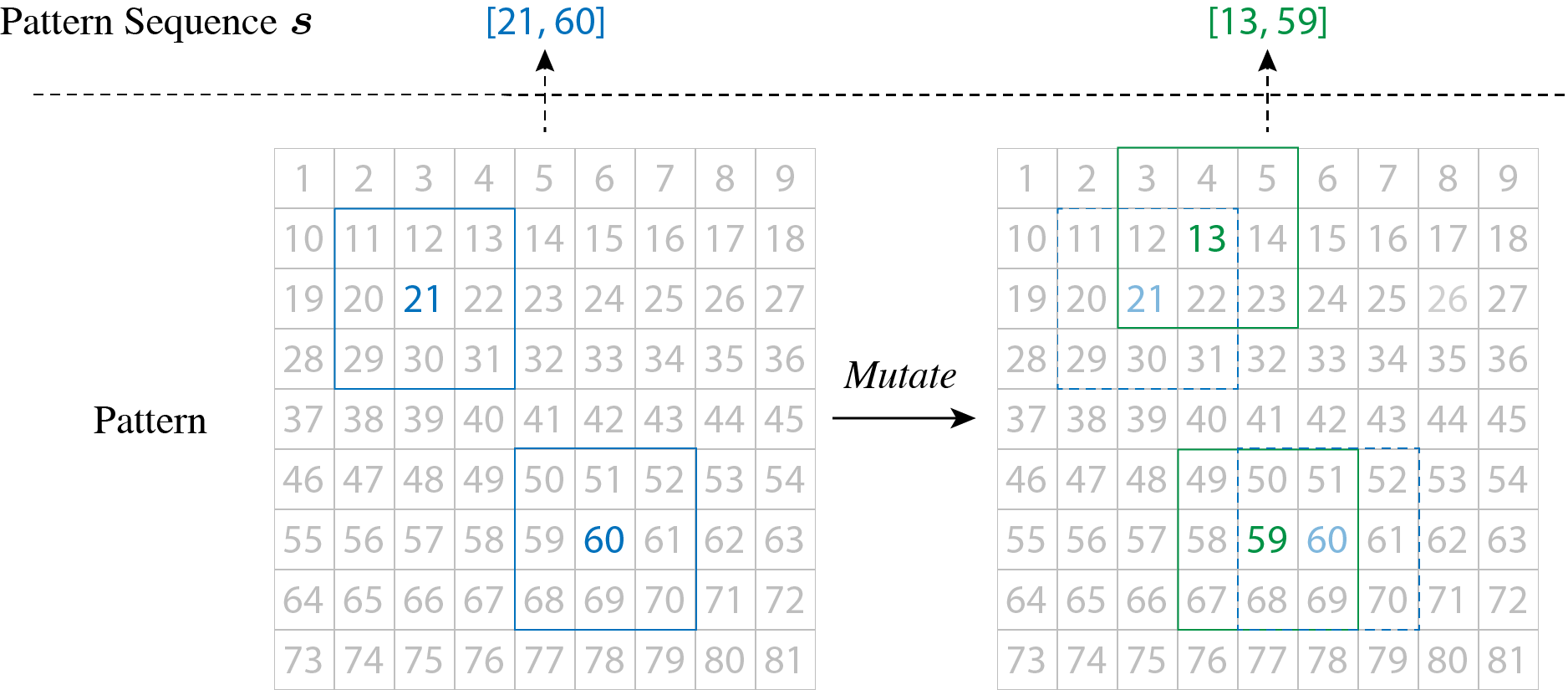}
\caption{An illustration of pattern mutation with 2 ABSs and a mutation rim of $r=1$. \vspace{-2ex}}
\label{fig:mutation}
\end{figure}

The mutation operation for our considered ABS pattern is illustrated in Fig. \ref{fig:mutation}, whereby a new ABS pattern (colored by green) is mutated from the initial ABS pattern (colored by blue).
We use the \textit{mutation rim} $r$ to control how far an ABS can mutate from its initial position, with $r=0$ representing no mutation, $r=1$ representing mutation to a one-hop grid, and so on.
As illustrated in Fig. \ref{fig:mutation},
the potential mutation range of each ABS is roughly bounded by its surrounding colored square, while the validity of a specific mutation position is rechecked by the ABS movement constraints \eqref{eq:moving_speed} - \eqref{eq:area}.
For the convenience of indexing and illustration, we introduce a \textit{pattern sequence} consisting of the flattened indexes\footnote{\rev{The 2D index in the $i$-th row and $j$-th column of a $K$-by-$K$ grid is denoted as $(i, j)$, while the corresponding one-dimensional (1D) flattened index is $(i-1)*K+j$. For example, in the left part of Fig. \ref{fig:mutation} with $K=9$, two ABSs locate at grids $(3, 3)$ and $(7, 6)$, with their flattened indexes given by $(3-1)*9+3=21$ and $(7-1)*9+6=60$, respectively.}} of residing grids for all ABSs, denoted by a vector $\boldsymbol{s}\in \mathbb{Z}^N$.
In addition, we eliminate patterns with overlapping ABSs during mutation since it is typically a waste to place two ABSs in the same grid.
Finally, for each period, we apply cKMeans as the initiating algorithm to find a base pattern of ABS locations.  

\subsubsection{Emulator-based Naive Mutation}
We first introduce a baseline version of mutation strategy called Naive Mutation (NM).
Specifically, from the initial base pattern, the NM strategy randomly and independently generates a certain number of mutated ABS patterns within the mutation range and also subject to the constraints \eqref{eq:moving_speed} - \eqref{eq:area}.
In the case without emulator assistance, the fitness evaluation of the mutated ABS patterns can only be performed \textit{physically} by on-site measurements within the limited time of each period. As a result, the NM strategy without emulator assistance can only try out a very limited number of mutated ABS patterns and thus has a low chance of achieving high CR.
Fortunately, with the help of our designed emulator,
the fitness evaluation of a certain ABS pattern can be performed \textit{virtually} by enquiring the emulator about its predicted CGU pattern and hence associated CR. Such emulator-based NM (termed \textit{SDL-NM}) enables us to virtually explore a much larger part of the search space (e.g., $N_\textrm{m}$ mutation attempts per period), under given budget of time and computational resource, thus leading to potential performance improvement.

However, the prediction errors due to emulator model approximation could also result in performance degradation.
To this end, we propose a simple mechanism termed as \textit{top-$k$ candidate election and validation}, or top-$k$ mechanism in short. The top-$k$ mechanism first elects $k$ top-performing candidate ABS patterns in terms of emulator-predicted CR, which are then physically tested by on-site measurements on their actual CR performance.
On one hand, such a mechanism eases the practical measurement trials since only a few most promising candidate solutions need to be verified. 
On the other hand, the on-site validations effectively compensate for the adversarial effect of emulator approximation errors, since top-$k$ candidates provide more robustness than top-one only.

Finally, despite of the emulator assistance and the top-$k$ mechanism, the performance of SDL-NM is still limited due to the following reasons.
First, all mutations of SDL-NM are originated from the same base pattern and thereby likely to trap in a local region of the search space.
Second, no iterative process is performed within the SDL-NM algorithm when generating new mutations.
In other words, there is no evolutionary progression since the mutated individuals are considered as peers and belong to the same generation.
The above defects hinder SDL-NM from finding better solutions, especially as the number of ABSs $N$ increases.

\subsubsection{Emulator-based MAP-Elites}

To circumvent the defects of SDL-NM, our proposed SDL-ME approach adopts another mutation strategy for ABS patterns based on MAP-Elites \cite{2015-MAP-Elites}, which belongs a recently emerged category of EAs called Quality-Diversity Algorithm (QDA) \cite{2016-Pugh-QD}.
Different from conventional EAs that aim for direct objective optimization, QDAs are more resemblant to the natural evolution process, which searches locally within each (ecological) niche by mutation as it simultaneously diversifies, thus promoting solution diversity for further achieving global optimality.

In particular, 
MAP-Elites first marks each candidate solution using a user-designated MD feature, and then keeps an archive of different (in terms of feature) solutions in that feature space, where each point (hereon termed \textit{niche}) in the feature space only stores the best-performing solution (in terms of objective) that maps to that niche. In other words, each niche is updated whenever a better-performing solution emerges with the feature pertaining to that niche.
As a result, the mutations from a base pattern gradually diversify in feature and simultaneously proliferate over a wide span of niches instead of trapping in a local region of the original search space.

In our considered scenario, we choose for simplicity two features to define the \textit{feature space}, i.e., the mean and standard deviation of the Euclidean distances between all ABSs, denoted by a two-dimensional (2D) vector $\boldsymbol b$, which take quantized values in 2D lattices or niches.\footnote{Other types of features can also be considered with proper trade-offs between solution diversity and algorithm complexity.}
Denote $\boldsymbol b=\Xi(\boldsymbol s)$ as the mapping function from the pattern sequence $\boldsymbol s$ to its corresponding feature niche $\boldsymbol b$.
In addition, denote $\lambda=\digamma(\boldsymbol s)$ as the fitness function that evaluates the CR $\lambda$ by enquiring the emulator with the ABS pattern corresponding to the pattern sequence $\boldsymbol s$.
Denote $\mathcal{B}$, $\Lambda$ and $\mathcal{S}$ as the sets that store the \textit{feature niche} $\boldsymbol b$, the \textit{highest recorded CR} $\lambda$ pertaining to $\boldsymbol b$, and the corresponding \textit{pattern sequence} $\boldsymbol s$ that achieves $\lambda$, respectively. With slight abuse of notations, for each feature niche $\boldsymbol b$, we can fetch its recorded CR and associated pattern sequence by $\lambda=\Lambda(\boldsymbol b)$ and $\boldsymbol s=\mathcal{S}(\boldsymbol b)$, respectively.
The pseudo code of our proposed SDL-ME algorithm is described in Algorithm \ref{alg:E_ME}.

The algorithm starts by initializing an empty archive $\Upsilon\triangleq\langle\mathcal{B}, \Lambda,\mathcal{S}\rangle$ that keeps track of high-performing solutions in the feature space (Line 1).
The base ABS locations are computed by the cKMeans algorithm in Section \ref{DataCollection} and then converted into a base pattern sequence $\boldsymbol{s}_0$ (Lines 2$\sim$3).
Next, a \textit{population generating process} is repeated for $N_{\text{it}}$ iterations each with $N_{\text{in}}$ individuals, thus generating a total number of $N_\textrm{m}\triangleq N_{\text{it}}N_{\text{in}}$ mutation attempts (Lines 4$\sim$18).
In each iteration, a pattern sequence set $\mathcal{S}^\prime$ is generated by either bootstrapping from $\boldsymbol{s}_0$ if the archive is empty, or randomly selecting $N_{\text{in}}$ pattern sequences from $\mathcal{S}$ to mutate (Lines 5$\sim$9). 
The mutated set $\mathcal{S}^{\prime}$ is then filtered by the movement constraints \eqref{eq:moving_speed} - \eqref{eq:area},
before mapping to the feature niche set $\mathcal{B}'$ and feeding to the emulator to get the corresponding fitness set $\Lambda'$ (Lines 10$\sim$12).
The mutated results in $\langle\mathcal{B}', \Lambda',\mathcal{S}'\rangle$ are then compared with the existing archive $\Upsilon$ to include a new entry $\langle\boldsymbol{b}, \lambda,\boldsymbol{s}\rangle$ or replace an existing entry by the one with higher CR (Lines 13$\sim$18). 
Finally, after $N_\text{it}$ rounds of iterations, the CR results in $\Lambda$ are sorted by descending order, yielding $k$ entries with emulator-believed top CRs. The best-$k$ pattern sequence set $\mathcal{S}^{*}\triangleq\{s^{*}_{1},\dots,s^{*}_{k}\}$ and hence the corresponding ABS location sets $\{\mathcal{P}^{*}_{1}, \dots, \mathcal{P}^{*}_{k}\}$ are then obtained (Lines 20$\sim$21).\footnote{\rev{The main computational workload in Algorithm \ref{alg:E_ME} lies in the emulator enquiry operation (feedforward NN inference) in Line 12, which is executed for a total number of $N_\textrm{m}\triangleq N_{\text{it}}N_{\text{in}}$ times.}}

\begin{algorithm}[t]\small
    \caption{Emulator-based MAP-Elites (E-ME)}\label{alg:E_ME}
    {{
        \KwIn{GU location set $\mathcal{Q}$ and initial ABS location set $\mathcal{P}$.}
        \KwOut{Emulator-believed best-$k$ ABS location sets $\{\mathcal{P}^{*}_{1}, \dots, \mathcal{P}^{*}_{k}\}$.}
        
        \SetKwFunction{FcKmeans}{cKMeans}
        
        Initialize an empty archive $\Upsilon\leftarrow\langle\mathcal{B}\leftarrow\emptyset, \Lambda\leftarrow\emptyset,\mathcal{S}\leftarrow\emptyset\rangle$\;
        Compute the base ABS location set $\mathcal{P}_{0}\leftarrow$ \FcKmeans($\mathcal{P}$, $\mathcal{Q}$)\;
        Convert $\mathcal{P}_{0}$ into pattern sequence $\boldsymbol{s}_{0}$\;
        
        \For{$i=1,\dots,N_{\text{it}}$}{
            \eIf{$i =1$}{
                Generate $N_{\text{in}}$ independent mutations based on $\boldsymbol{s}_{0}$ to bootstrap a pattern sequence set $\mathcal{S}^{\prime}$\;
            }{
                Randomly select $N_{\text{in}}$ pattern sequences from $\mathcal{S}$ to mutate, and get a pattern sequence set $\mathcal{S}^{\prime}$\;
            }
            Filter $\mathcal{S}^{\prime}$ by removing $\boldsymbol{s}$ that does not satisfy the constraints \eqref{eq:moving_speed} - \eqref{eq:area}\;
            Feed $\mathcal{S}^{\prime}$ to the mapping function to get the feature niche set $\mathcal{B}^{\prime}\leftarrow\Xi(\mathcal{S}^{\prime})$\;
            Feed $\mathcal{S}^{\prime}$ to the emulator to get the fitness set $\Lambda'\leftarrow\digamma(\mathcal{S}^{\prime})$\;
            
            \For{each $\langle\boldsymbol{b}, \lambda,\boldsymbol{s}\rangle\in\langle\mathcal{B}', \Lambda',\mathcal{S}'\rangle$}{
                \If{$\boldsymbol{b}\not\in\mathcal{B}$ or $\Lambda(\boldsymbol{b}) < \lambda$}{
                    $\mathcal{B}\leftarrow \boldsymbol{b}$; $\Lambda(\boldsymbol{b}) \leftarrow \lambda$; $\mathcal{S}(\boldsymbol{b}) \leftarrow \boldsymbol{s}$\;              
                }
            }
        }
        Get emulator-believed best-$k$ pattern sequence set $\mathcal{S}^{*}\triangleq\{s^{*}_{1},\dots,s^{*}_{k}\}$ from $\Upsilon$ according to $\Lambda$\;
        Convert $\mathcal{S}^{*}$ to ABS location sets $\{\mathcal{P}^{*}_{1}, \dots, \mathcal{P}^{*}_{k}\}$.
    }}
\end{algorithm}

\begin{remark}
 Our proposed SDL-ME algorithm features the following characteristics.
(1) The search over the original variable space (regarding a $K$-by-$K$ matrix $P_p$) is effectively reduced to the search over a \textit{customized low-dimensional feature (or observation) space}, which allows flexible trade-off between complexity reduction and solution diversity.
(2) Thanks to the archive structure, the search for a solution in any single niche is aided by the simultaneous search for solutions in other niches. This brings \textit{cross-generation progressions} and enables more extensive explorations in the feature space, which help avoid local optima and find different, often better, performance peaks.
(3) The archive structure and the iterative progression are supported by our proposed environment emulator, which allows \textit{virtual mutations} across different regions almost freely, without frequent and time-consuming trials and errors on site.
Besides the ease on the physical operation side, the emulator-based virtual planning significantly boosts the search progress by allowing \textit{much more attempts} under given time budget and computational resource. 
In summary, all the above characteristics of SDL-ME encourage more efficient search for globally better solutions. 
\end{remark}

\subsection{PES Scheme for Practical Multi-ABS Movement Operation}\label{PES}
Our emulator-based planning algorithms, i.e., E-NM and E-ME, both return \rev{their own} top-$k$ candidate solutions before the start of each period.
These recommended solutions are later deployed and evaluated on site in the exploration phase, out of which the best performing one is adopted in the serving phase.
In the following, we focus on the trajectory design for exploration and serving.

In the exploration phase, the top-$k$ candidate ABS location sets are evaluated on site based on a certain visiting order.
Since the multi-ABS movement coordination involves $N$ ABSs each with $k$ target waypoints, determining the optimal visiting order in terms of overall coverage and/or traversal distances could itself be a tough task.
Fortunately, our main objective in the exploration phase is to verify the CR of candidate solutions and to elect the best one for the serving phase which typically has a longer duration $\Delta t_{s}\ge\Delta t_{e}$. For simplicity, the $k$ candidate solutions are visited based on the descending order of their predicted CRs, while more advanced design is left for future work.

\begin{figure}
    \centering
    \begin{subfigure}[b]{0.4\linewidth}
        \centering
        \includegraphics[width=1\linewidth,  trim=40 40 40 40,clip]{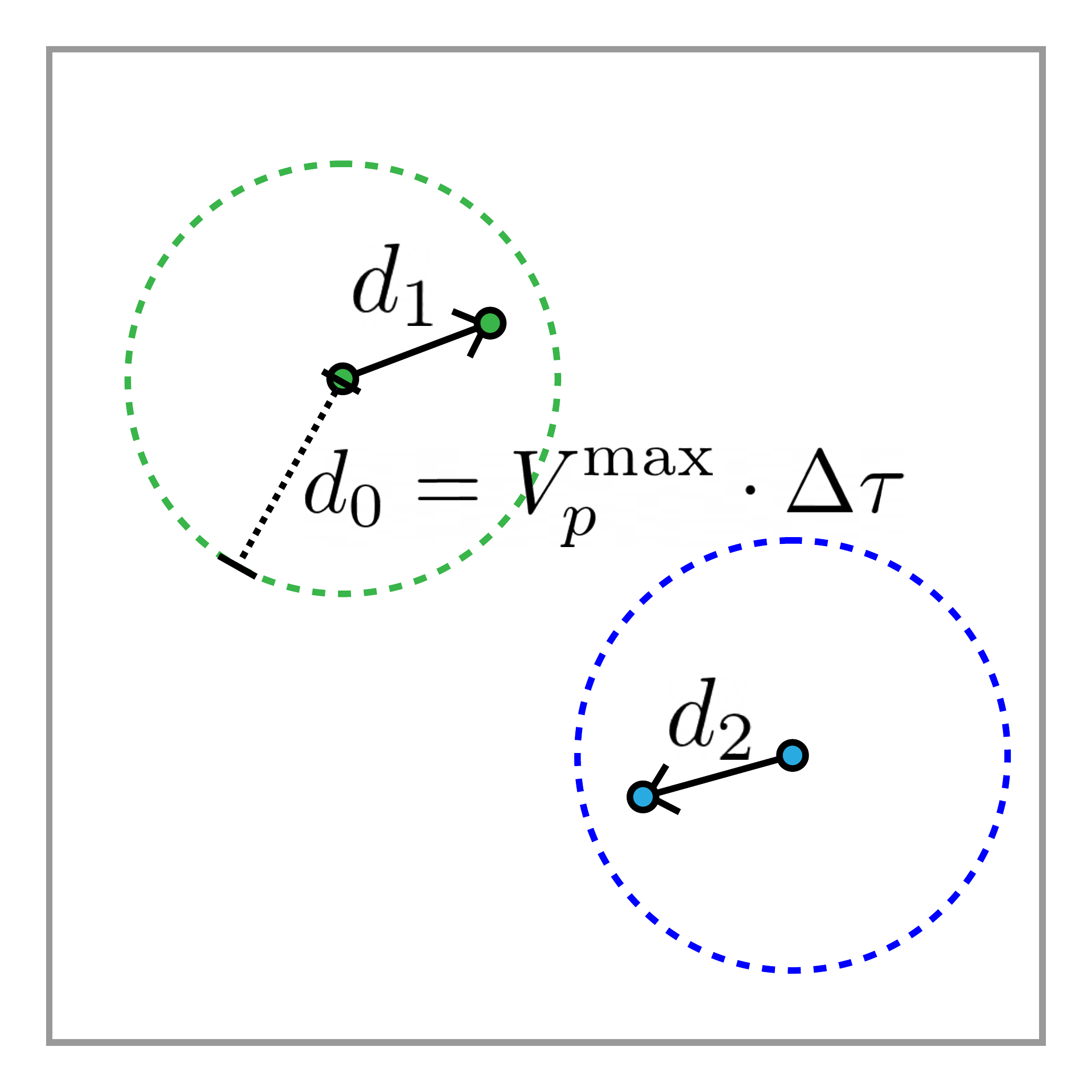}
        \caption{Case 1 \vspace{-1ex}}
        \label{fig:type1}
    \end{subfigure}
    \hspace{5ex}
    \begin{subfigure}[b]{0.4\linewidth}
        \centering
        \includegraphics[width=\linewidth,  trim=40 40 40 40,clip]{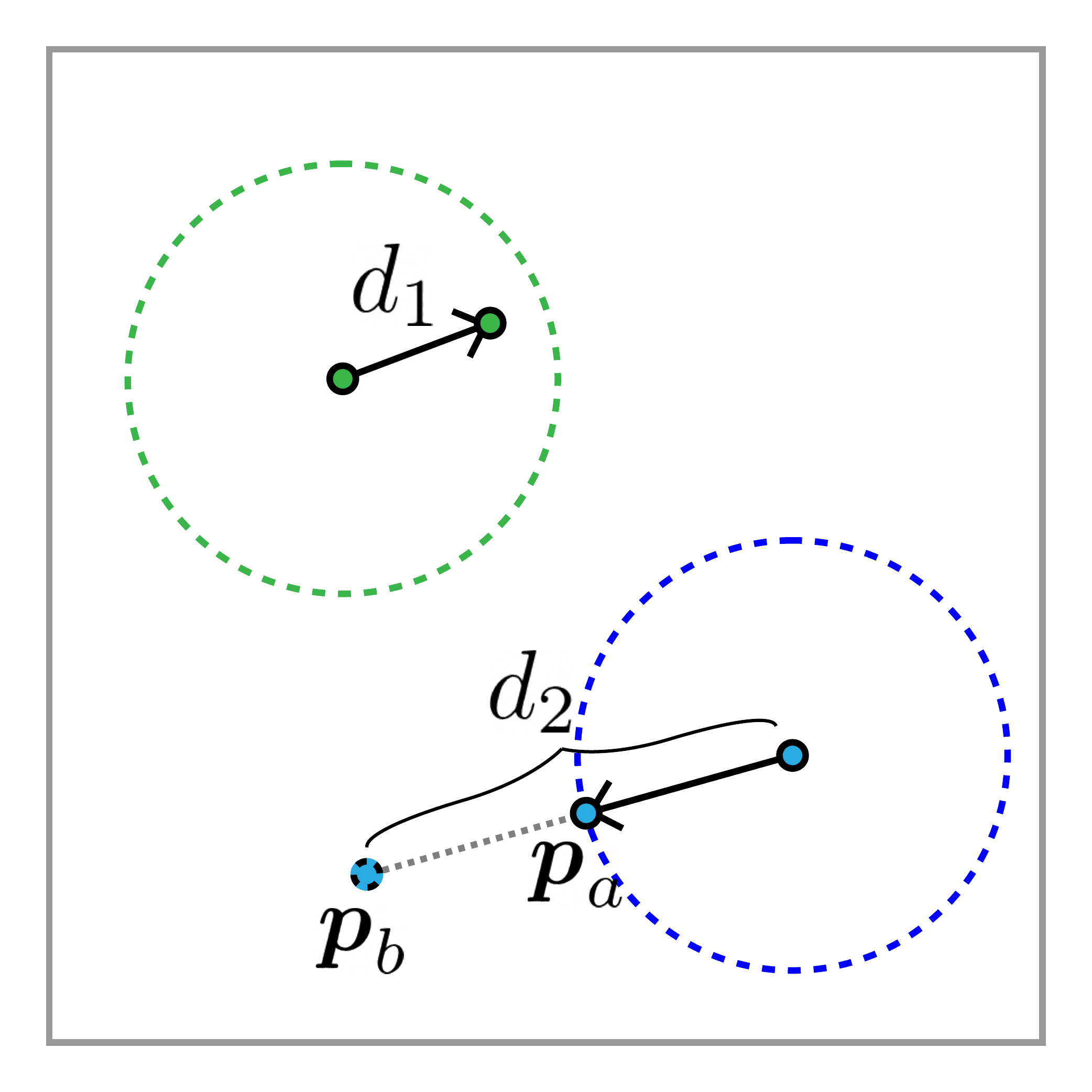}
        \caption{Case 2 \vspace{-1ex}}
        \label{fig:type2}
    \end{subfigure}
    \caption{An illustration of trajectory moving rules between two consecutive steps when $N=2$. \vspace{-1ex}}
    \label{fig:trajectory}
\end{figure}

Another practical issue is how to deal with the non-uniform flight distances between two adjacent waypoints for all ABSs to move coordinately.
Since the ABS-GU connectivity measurement is performed on a step basis and the ABS movement is constrained by a maximum moving speed $V^{\max}_{p}$, it could happen that an ABS cannot fly to its target location within a step time $\Delta \tau$.
For illustration, two typical cases are depicted in Fig. \ref{fig:trajectory} with 2 ABSs.
The maximum displacement of an ABS between two consecutive steps is bounded by $d_0 \triangleq V^{\max}_{p}\cdot\Delta \tau$.
In the first case, as shown in Fig. \ref{fig:type1}, the target locations for both green and blue ABSs are within the valid displacement range of one step, i.e., $d_1 \leq d_0$ and $d_2 \leq d_0$.
In this case, each ABS can reach its target location within a single step, and we assume that all ABSs adjust their speeds so that they can arrive at their target locations simultaneously at the end of the time step.
In the second case, there exists at least one ABS whose distance from its target location is large than the valid range $d_0$, e.g., the blue ABS with distance $d_2$ from its target location $\boldsymbol p_b$, as shown in Fig. \ref{fig:type2}.
In this case, we assume that the blue ABS flies towards its target location at maximum speed $V^{\max}_p$, and arrives at an intermediate location $\boldsymbol p_a$ at the end of the current step. In the upcoming steps, the blue ABS continues moving towards $\boldsymbol p_b$ at maximum speed until reaching it, while other ABSs (e.g., the green ABS) hover at their arrived target locations. Connectivity measurements are still performed regularly on a step basis during the above process.
Finally, after all $k$ candidate solutions are traversed or the exploration time $\Delta t_{e}$ runs up, the per-step measurements in the exploration phase elect the best recorded solution $\mathcal{P}^*$ with the highest CR. All ABSs then fly at maximum speed towards their corresponding locations in $\mathcal{P}^*$, respectively, and serve there till the end of the current period.

\begin{remark}
    The amalgamation of emulator-based planning and on-site exploration/serving significantly reduces the extensive step-by-step on-site trials and errors. In addition, the on-site measurements effectively compensate for the quantization/prediction errors due to model approximation.
\end{remark}


\section{Numerical Results}
\label{sec:simulation_result}
In this section, we present comprehensive simulation results to demonstrate the effectiveness of our proposed SDL-ME approach.
Comparisons are made against one of the state-of-the-art DRL methods.
In addition, we include Naive Mutation with constrained K-means algorithm (NM) (without emulator assistance) as the very basic baseline, as well as its model-based version, i.e., SDL-NM.
The performance metric is either the ACR of the entire trial, or the CR of each step.
The default parameters are listed in the following if not stated otherwise:
$D=1000$ m, $D_w=31.25$ m, $h_w[l] \sim \textrm{Uniform}(30,89)$ m, $L=200$, $K=64$, $N=5$, $M=100$, $d_{\min}=10$ m, $h_{p}=60$ m, $h_{q}=1$ m, $V^{\max}_p=30$ m/s, $V_q=2$ m/s, $\Delta T=200$ s, $\Delta t=10$ s, $\Delta t_e=5$ s, $\Delta t_s=5$ s, $\Delta t_p=3$ s, $\Delta \tau=0.5$ s, 
$k=10$, $r=3$, $N_{\text{m}}=8192$, $N_{\text{it}}=64$, $N_{\text{in}}=128$,
\rev{$K_{\textrm{min}}=0$ dB, $K_{\textrm{max}}=30$ dB, carrier frequency $f_\textrm{c}=2$ GHz, $B=20$ MHz, and $\gamma_{\text{t}}=115$ dB.
To better utilize the limited resources per ABS and share the serving task, the maximum number of supported GUs per ABS is configured as $M_\textrm{max}=\lfloor(1+\epsilon) M/N\rfloor$, where $\epsilon\geq 0$ is a constant (e.g., 0.2).
}
The simulations are conducted using PyTorch 1.10 with 16-core 3.8 GHz CPU, 64 GB memory, and one Nvidia RTX 2080 Ti GPU.

\subsection{SDL-ME versus DRL}
\subsubsection{Benchmark DRL Method}
\rev{We adapt the algorithm proposed in \cite{2020-Lyu-Codesign} to fit our coverage problem, and replace the underlying DDPG framework with its advanced variant --- Twin Delayed Deep Deterministic policy gradient algorithm (TD3) framework \cite{2018-TD3}.
Compared to DDPG, TD3 has three additional tricks that enable faster convergence and better performance, including clipped double Q-learning, delayed policy updates and target policy smoothing, which are duly implemented here.
Since the original paper \cite{2020-Lyu-Codesign} only considers a dual-ABS scenario with $10$ GUs, here we consider $N=2$ ABSs serving $M=20$ GUs with a target throughput of $\bar R=3.6$ Mbps per GU in an environment with site-specific blockages\footnote{\rev{For illustration, an urban region map in Beijing is used here (see Fig. \ref{fig:demo}), which is imported from paid dataset such as Baidu Map.}} for method-to-method performance comparison.

For the benchmark TD3 method, the state comprises the 2D positions of all ABSs and GUs, which is a vector of $(2+20)\times 2=44$ scalar values in this example setup.
The action comprises the $x$-axis and $y$-axis moving distances for all ABSs, which is a vector of $2\times 2$ scalar values.
Finally, the reward function is based on the instantaneous coverage rate added by a penalty (set to $-100$) if any of ABSs moves out of the considered area.
As for the NN architecture, the actor consists of $3$ fully connected layers each with $256$ neurons, where the first two layers are each followed by a rectified linear unit (ReLU) activation function while the last layer is followed by a hyperbolic tangent (Tanh) activation function.
The critic has two copies of $3$ fully connected layers each with $256$ neurons, where the first two layers are also followed by ReLU activations.
The learning rates of the actor and critic are both $0.0003$;
the discount factor is $0.995$; 
the standard deviation of Gaussian exploration noise is $0.25$; 
the noise added to target actor during critic update is $0.2$; 
the noise clipping boundary is $0.5$; 
the delayed policy update frequency is $4$; 
the buffer size of replay memory is set as $10^6$ and the data is sampled uniformly.
The above hyperparameters are set with reference to \cite{2020-Lyu-Codesign} and also tuned with best effort for performance improvement.
}

\subsubsection{Performance Comparisons}
For TD3, the agent is trained with a batch size equal to $256$ for $20,000$ trials with $400$ steps in each trial. The offline training time is about 32 hours for TD3 to converge.
In contrast, our SDL-ME algorithm only requires samples collected from $2,400$ trials with $400$ steps in each trial, while the corresponding offline training only takes about 5 hours on the same computer for the emulator to achieve desired model accuracy. Therefore, our proposed SDL-ME method outperforms TD3 in terms of both \textit{sample/training efficiency}.

\begin{figure*}[!t]
	\centering
	\begin{subfigure}[b]{0.32\textwidth}   
		\centering 
		\includegraphics[width=\textwidth,  trim=0 50 0 50,clip]{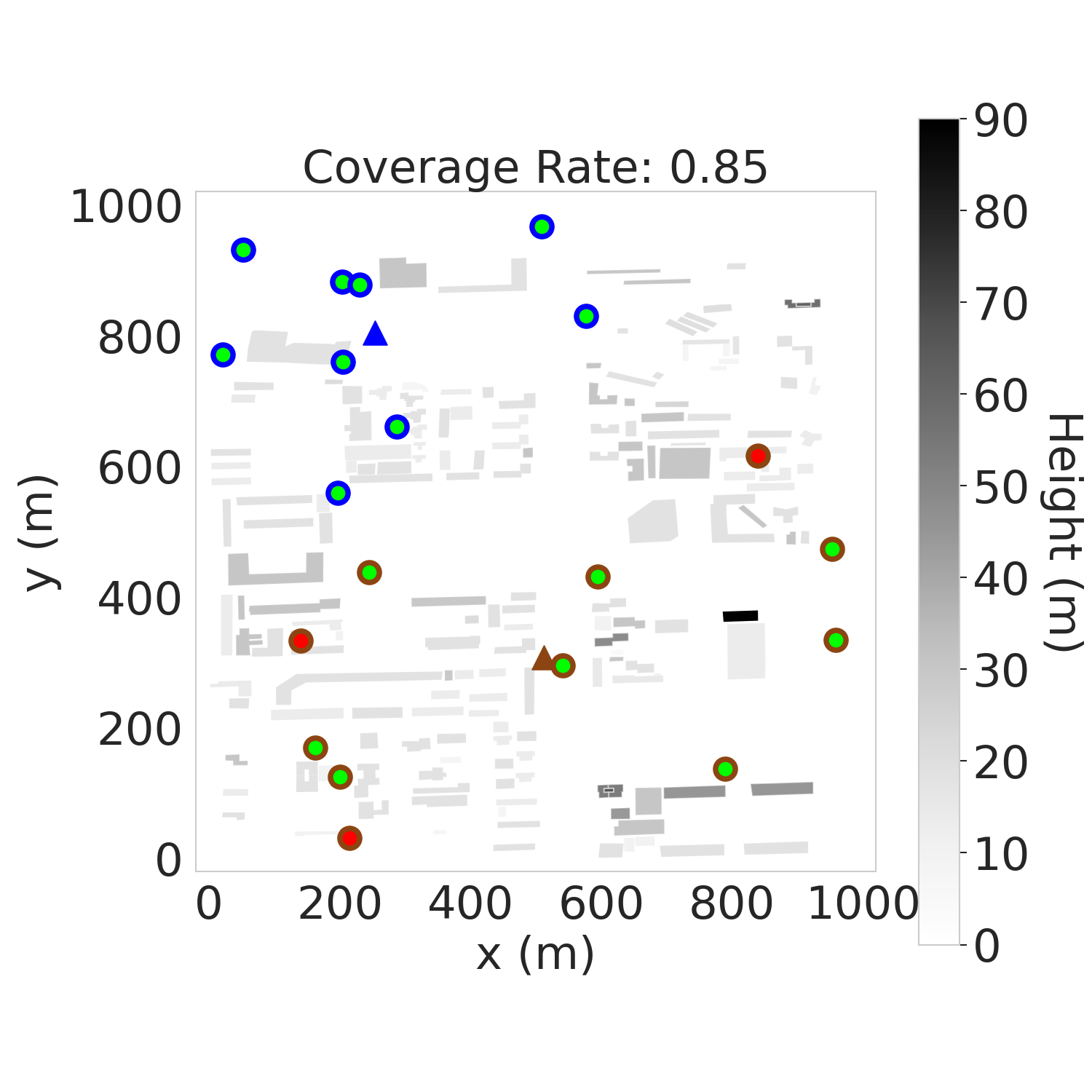}
		\caption[Initial Status]%
		{{\small Initial Status}}\vspace{-2ex}  
		\label{fig:demo_init}
	\end{subfigure}
	\hfill
	\begin{subfigure}[b]{0.32\textwidth}   
		\centering 
		\includegraphics[width=\textwidth,  trim=0 50 0 50,clip]{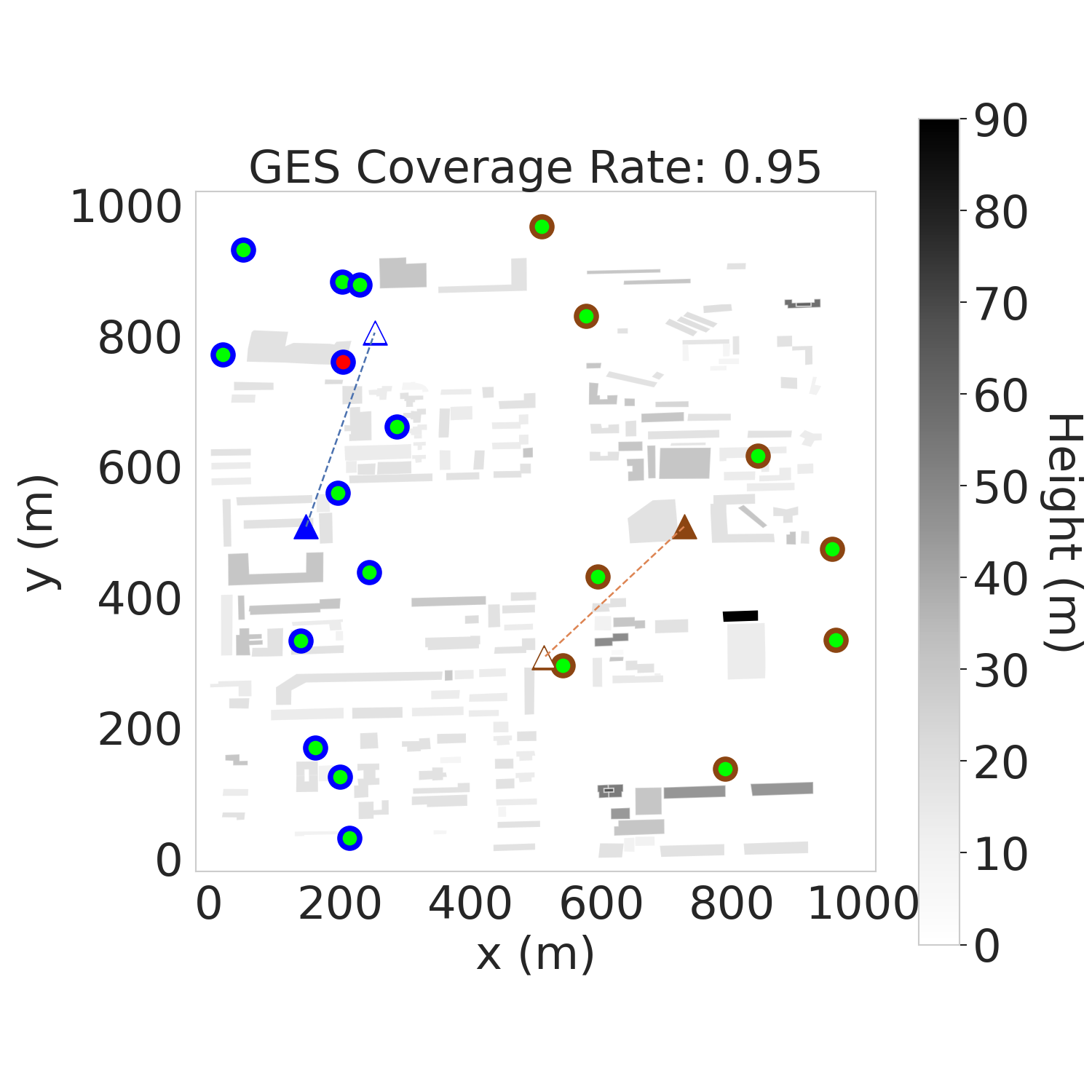}
		\caption[GES]%
		{{\small GES}}\vspace{-2ex}    
		\label{fig:demo_ES}
	\end{subfigure}
	\hfill
	\begin{subfigure}[b]{0.32\textwidth}  
		\centering 
		\includegraphics[width=\textwidth,  trim=0 50 0 50,clip]{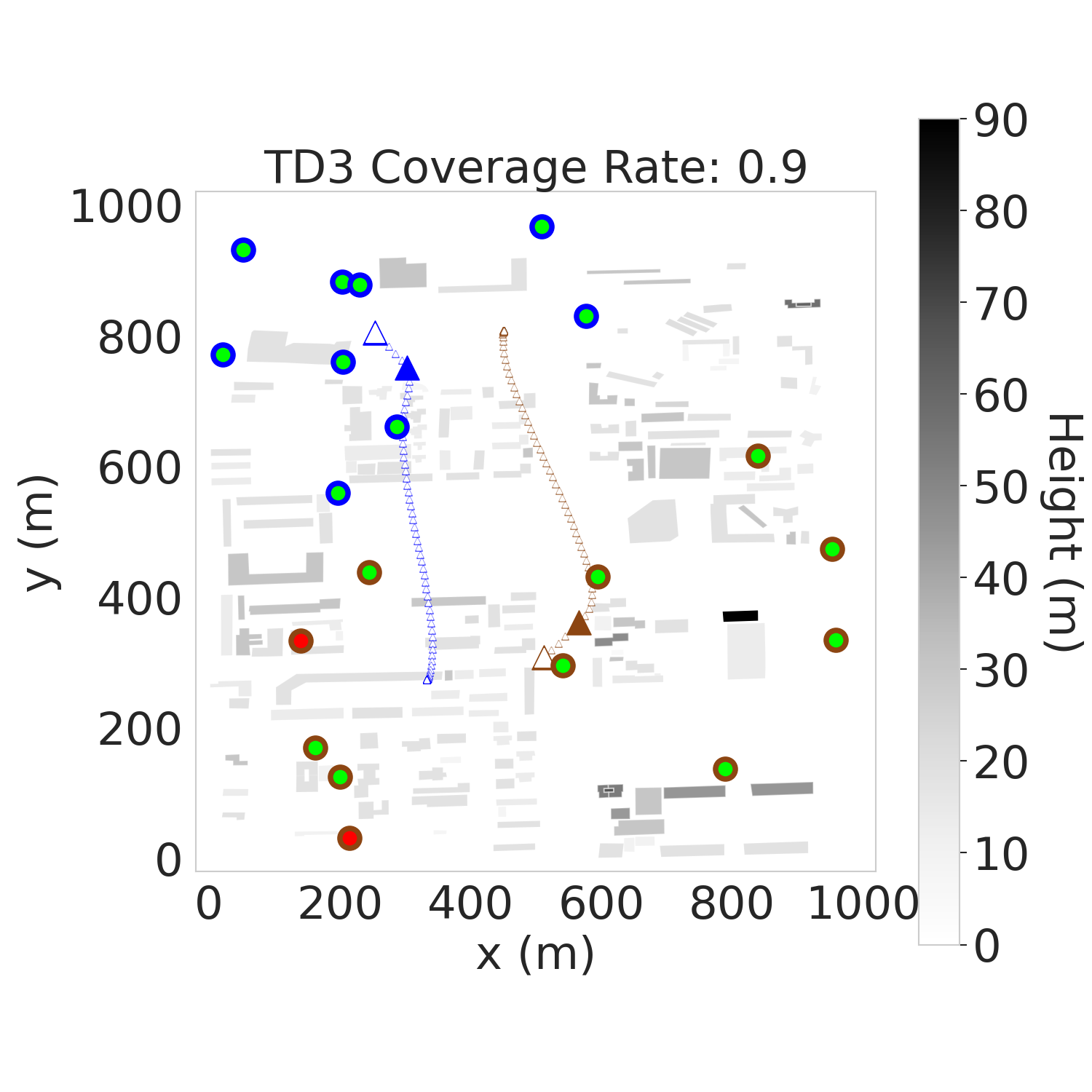}
		\caption[TD3]%
		{{\small TD3}}\vspace{-2ex}    
		\label{fig:demo_TD3}
	\end{subfigure}
	\vskip\baselineskip
	\begin{subfigure}[b]{0.32\textwidth}
		\centering
		\includegraphics[width=\textwidth,  trim=0 50 0 50,clip]{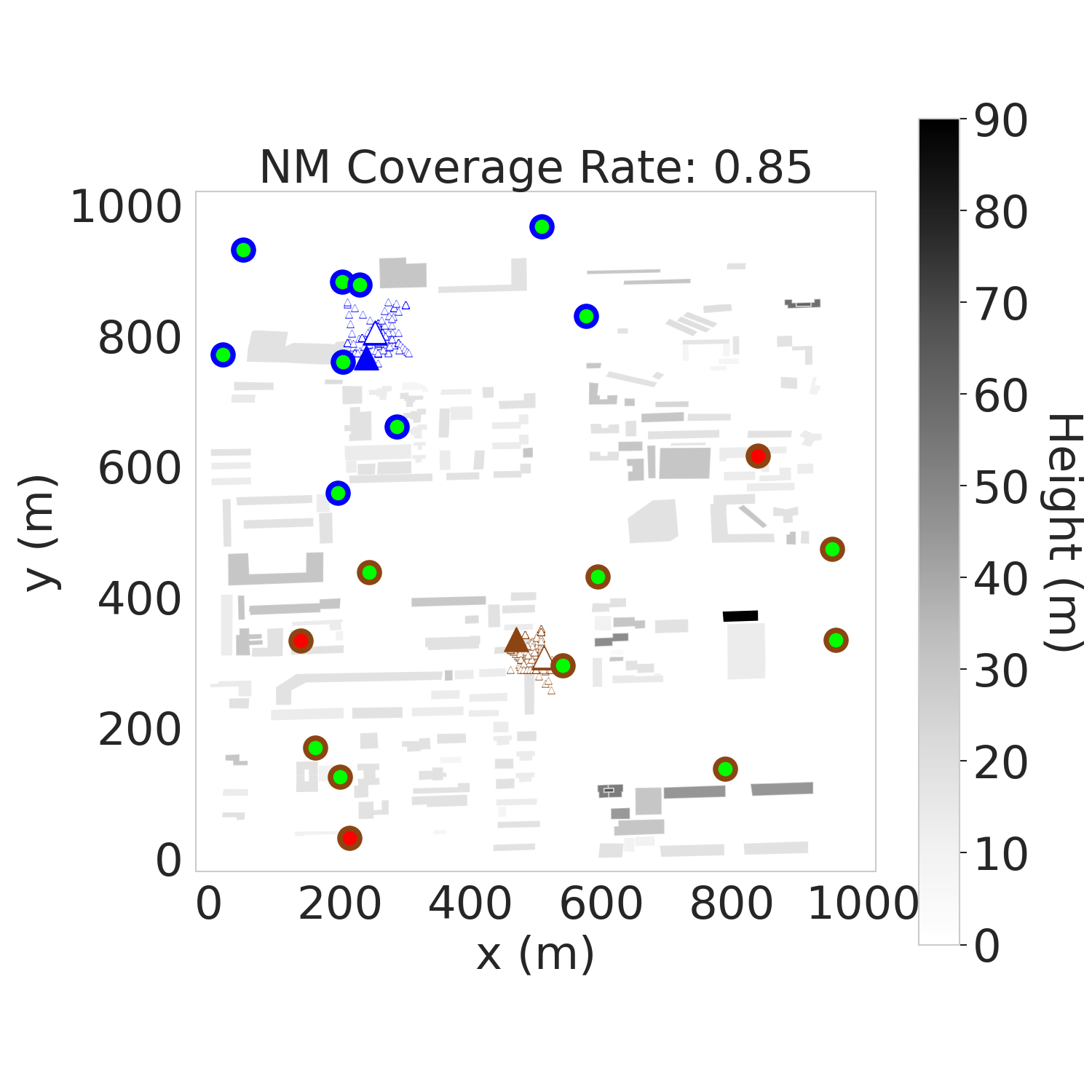}
		\caption[NM]%
		{{\small NM}}\vspace{-1ex}    
		\label{fig:demo_NM}
	\end{subfigure}
	\hfill
	\begin{subfigure}[b]{0.32\textwidth}   
		\centering 
		\includegraphics[width=\textwidth,  trim=0 50 0 50,clip]{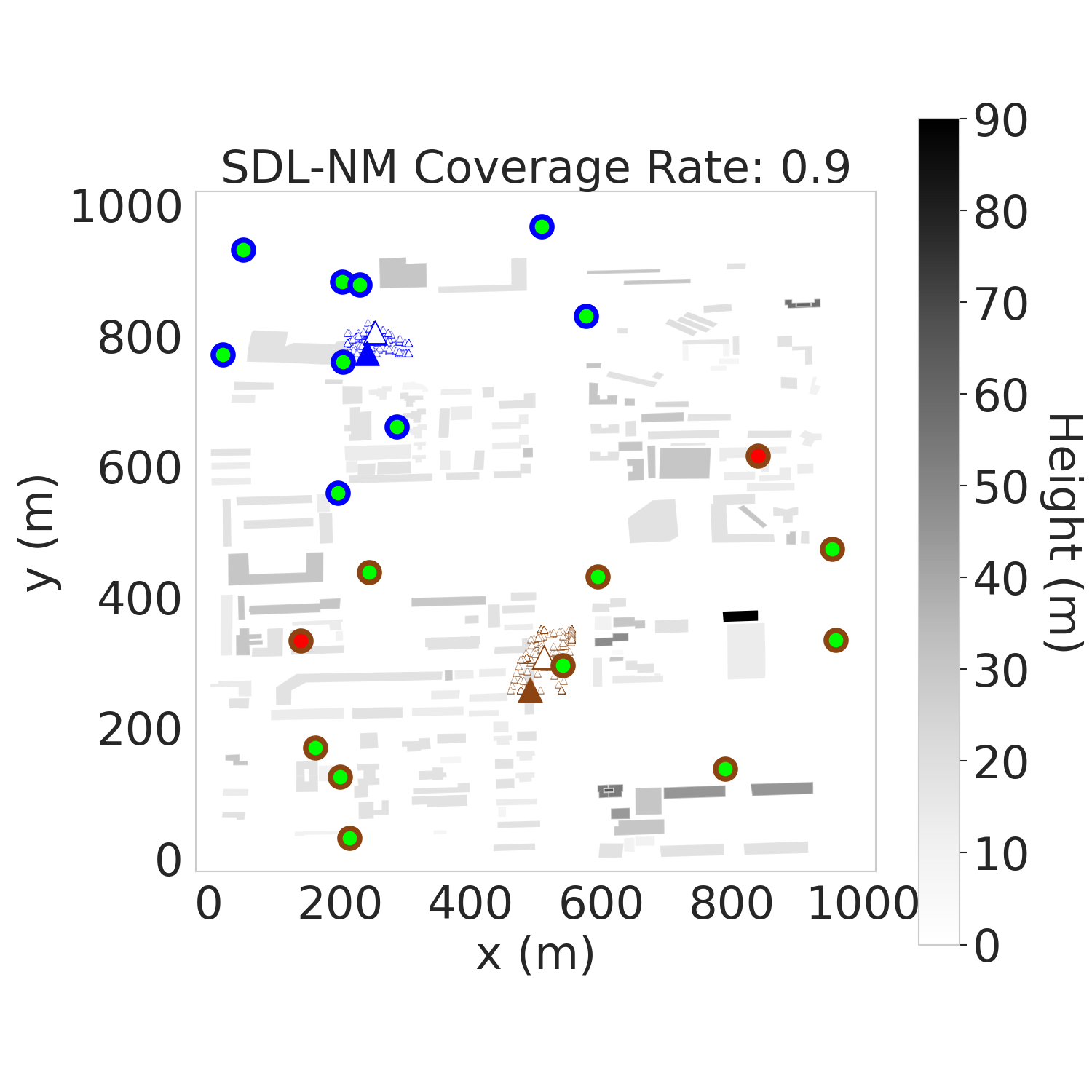}
		\caption[SDL-NM]%
		{{\small SDL-NM}}\vspace{-1ex}    
		\label{fig:demo_SDL_NM}
	\end{subfigure}
	\hfill
	\begin{subfigure}[b]{0.32\textwidth}   
		\centering 
		\includegraphics[width=\textwidth,  trim=0 50 0 50,clip]{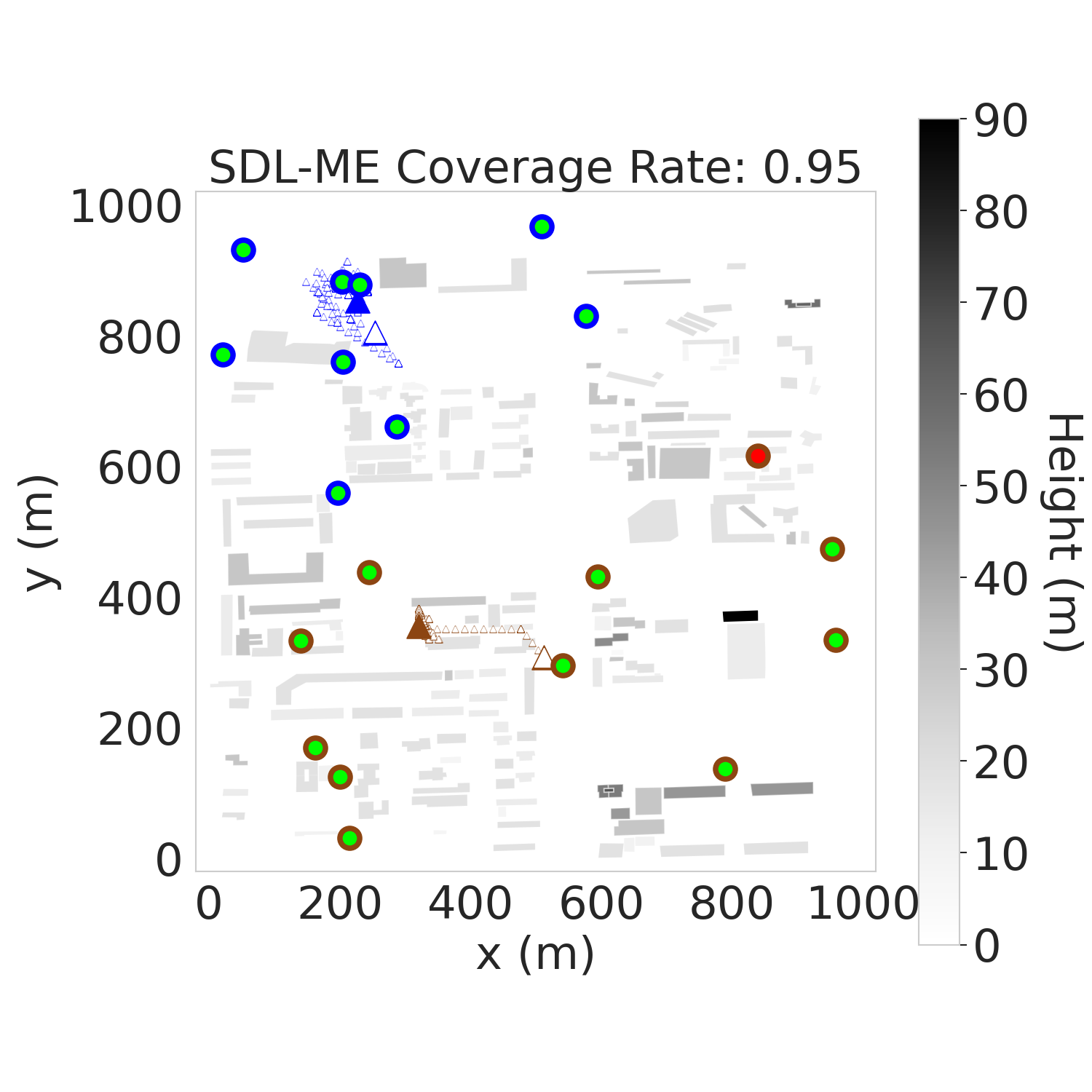}
		\caption[SDL-ME]%
		{{\small SDL-ME}}\vspace{-1ex}    
		\label{fig:demo_SDL_ME}
	\end{subfigure}
	\hfill
	\caption{\rev{ABS placement solution by different schemes under the same initial GU/ABS locations ($N=2$, $M=20$). The grey region represents the building area, whose color darkness indicates the building heights. Each ABS is represented by a solid triangle, with either blue or brown color, respectively. Each circle represents a GU, whose contour line is either blue or brown to indicate its associated ABS, and the filling is either green or red to indicate either covered or uncovered, respectively. The ABS movement trajectories are shown in dashed lines (empty triangles represent the initial ABS positions).}}\vspace{-2ex}
	\label{fig:demo}
\end{figure*}

\rev{For small-scale problems, we also obtain the performance upper bound by \textit{grid exhaustive search (GES)}, which tries out all possible combinations of ABS positions in the grids subject to the movement constraints \eqref{eq:moving_speed} - \eqref{eq:area},\footnote{\rev{The maximum ABS displacement within the exploration time $\Delta t_e$ is given by $V^{\max}_p\Delta t_e$, which translates into a search radius of $r_{\textrm{max}}\triangleq\frac{V^{\max}_p\Delta t_e}{D/K}$ grids per ABS. As a result, the total number of ABS patterns to try out in GES is $O\big((r_{\textrm{max}})^{2N}\big)$, which is exponential in $N$.}} and assumes that the optimal solution is found instantaneously at the start of each period.}
In Fig. \ref{fig:demo}, we first visualize the placement solution of different schemes for one period \textit{under given GU locations}.
The initial placement solution based on K-means is shown in Fig. \ref{fig:demo_init}. 
The performance upper bound is attained via GES, as shown in Fig. \ref{fig:demo_ES}.
The results for other schemes are shown from Fig. \ref{fig:demo_TD3} to Fig. \ref{fig:demo_SDL_ME}, respectively.

For the considered example setup, the CR ranking by ascending order for all schemes is given by (a) Initial Status, (d) NM, (c) TD3, (e) SDL-NM, (f) SDL-ME and (b) GES, where SDL-ME achieves the same highest CR compared with GES, and outperforms other schemes including TD3.
\rev{The ABS movement trajectories in one period are shown in dashed lines in Fig. \ref{fig:demo}. Note that for the TD3 method, it is typically difficult to find desirable solutions within a single period. As a result, we extend its exploration time from one period to one trial, and retain the best CR result along the path, as shown in Fig. \ref{fig:demo_TD3}.
On the other hand, the NM and SDL-NM schemes tend to search locally around the initial K-means solution, while the SDL-ME scheme might attempt to search in wider regions for better solutions.}
For the NM method, without the emulator assistance, it only tries out a small proportion of the search space with a very limited number of steps, whereby the resultant placement pattern is the most resemblant to the initial K-means pattern.
For SDL-NM and SDL-ME, they knock out most of the inferior placement patterns using emulator-based planning, and only perform the emulator-believed top candidate solutions to select the actual best pattern during each period.
Unlike SDL-NM that generates mutations from a single base pattern, SDL-ME keeps an archive of different patterns in a designed feature space and update the best seen patterns iteratively, which in turn is more likely to achieve better performance.

\begin{figure}[htbp]
	\centering
	\includegraphics[width=0.95\linewidth,  trim=0 0 0 0,clip]{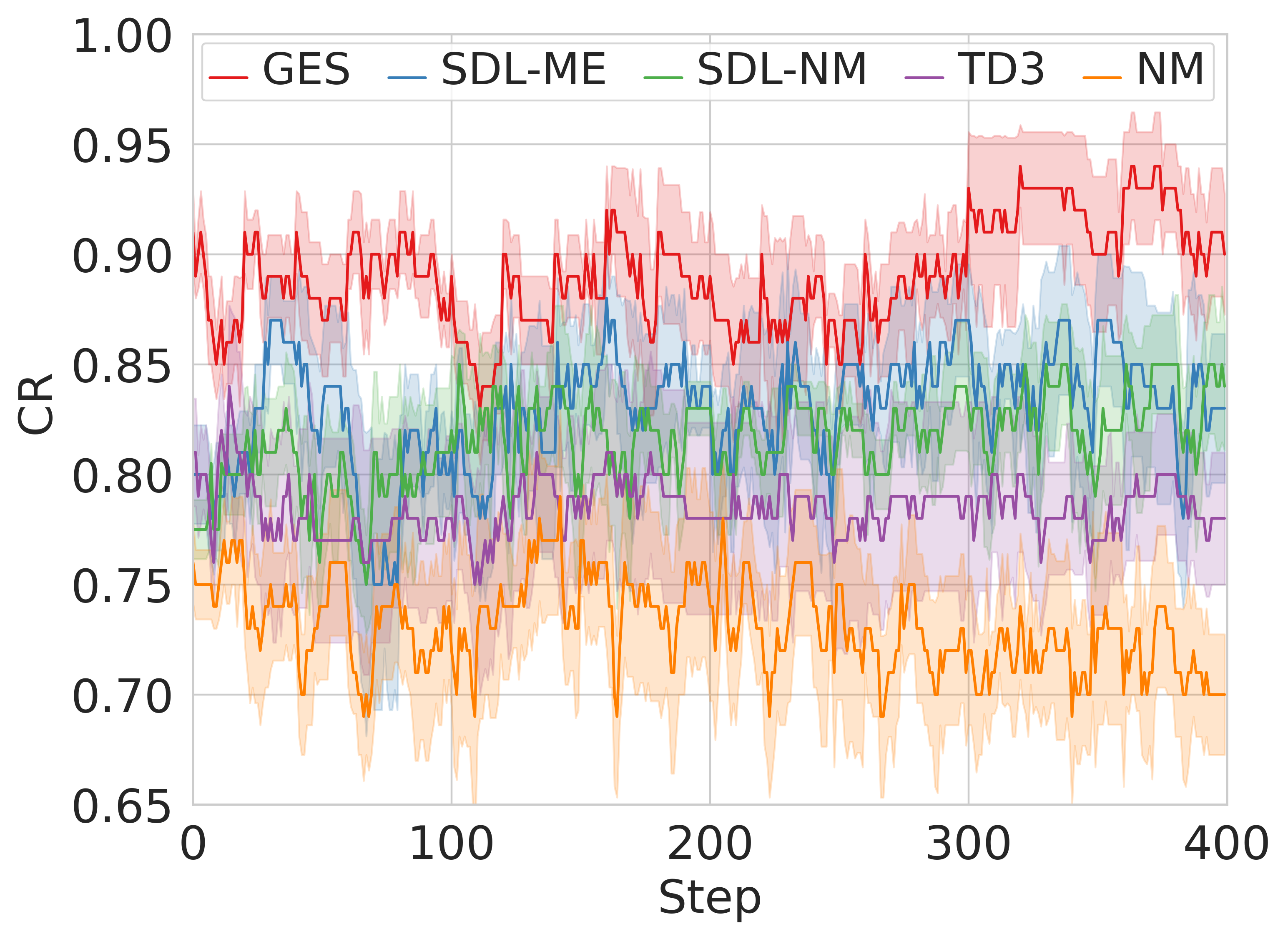}
	\caption[]{\rev{Step-wise CR performance of the four schemes under 5 different trials ($N=2$, $M=20$). For each scheme, the solid line represents the average CR per step while the corresponding shaded region represents the standard deviation.}}\vspace{-2ex}
	\label{fig:case_1_1}
\end{figure}

Next, we consider the \textit{dynamic case with moving GUs} as described in Section \ref{SectionMobility}, and evaluate the step-wise CR performance by averaging over $5$ different trials, each with $400$ steps.
The results are shown in Fig. \ref{fig:case_1_1}.
It can be seen that TD3, SDL-NM and SDL-ME all outperform the baseline NM method.
Notably, SDL-NM and SDL-ME also outperform TD3, thus advocating the advantage of our SDL-based method in terms of step-wise CR performance.
Finally, a performance gap between SDL-ME and GES is still observed in Fig. \ref{fig:case_1_1}.
\rev{Nevertheless, it is worth noting that the results obtained by GES only serve as a \textit{loose} upper bound, which is difficult, if not impossible, to achieve in practice. 
This is due to the facts that 1) in practice, each period has only $J$ (e.g., 20) steps (chances) for on-site connectivity measurements; 
2) the GES method assumes that the search (and the corresponding ABS movement actions) is completed \textit{instantaneously} at the start of each period whereas such computation/exploration has an exponential complexity in $N$, thus ignoring both the time and cost behind all these trials and errors.
}


\begin{figure}[htbp]
	\centering
	\includegraphics[width=0.85\linewidth,  trim=0 0 0 0,clip]{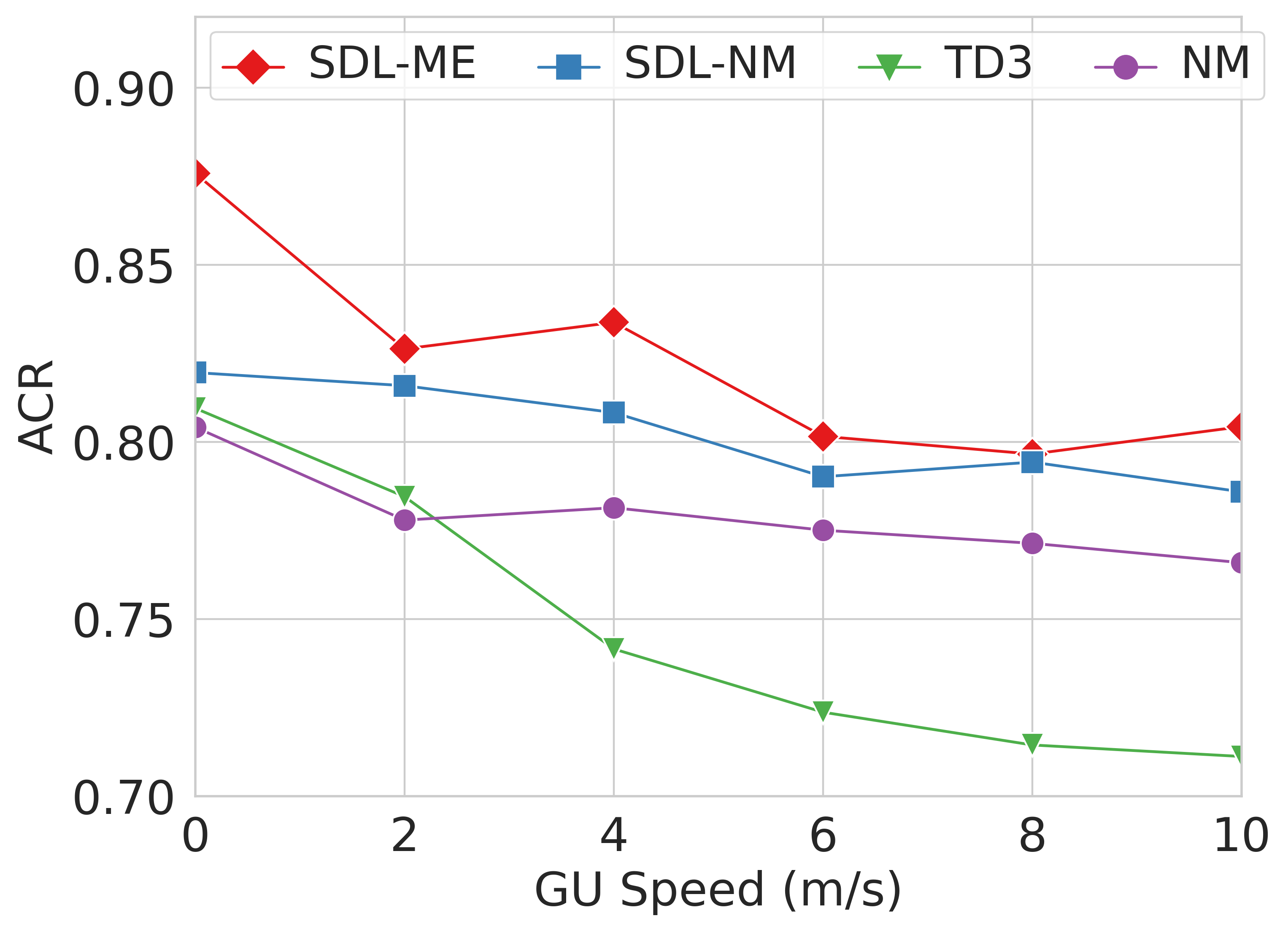}
	\caption[]{\rev{ACR performance and robustness of the four schemes under different GU speeds ($N=2$, $M=20$). The model parameters of SDL-ME, SDL-NM and TD3 are offline trained under $V_{q}=2$ m/s.}}\vspace{-2ex}
	\label{fig:case_1_2}
\end{figure}

Furthermore, in Fig. \ref{fig:case_1_2}, we compare the ACR performance and robustness of the four schemes under \textit{different GU speeds}, where SDL-ME, SDL-NM and TD3 are all offline trained under the default GU speed (i.e., $V_{q}=2$ m/s).
First, it can be seen that SDL-ME achieves remarkably higher ACR than those of other schemes under all the tested GU speeds.
Second, for TD3, the highest ACR is achieved under $V_{q}\leq 2$ m/s (the default GU speed $V_{q}=2$ m/s is used for offline training), whereas the ACR decreases sharply as the GU speed increases.
In comparison, degradation of ACR for SDL-ME is relatively slow, thus demonstrating robustness under small/moderate GU speeds. 
The underlying reasons could be explained as follows.
For TD3, it belongs to the DRL methods which inherently possess strong step-to-step correlations, and hence the policy learned during offline training becomes no longer suitable under a different evaluation condition (e.g., GU speed in this case).
On the other hand, for SDL-based approaches, there is no strong step-to-step correlations despite the maximum movement distance constraint in each time period.
As a result, SDL-based approaches show more flexibility and robustness compared with DRL-based methods under different GU speeds.

\subsection{Accommodation To Larger Problems with Complex/Dynamic Scenarios}
\begin{figure}[htbp]
	\centering
	\includegraphics[width=0.85\linewidth]{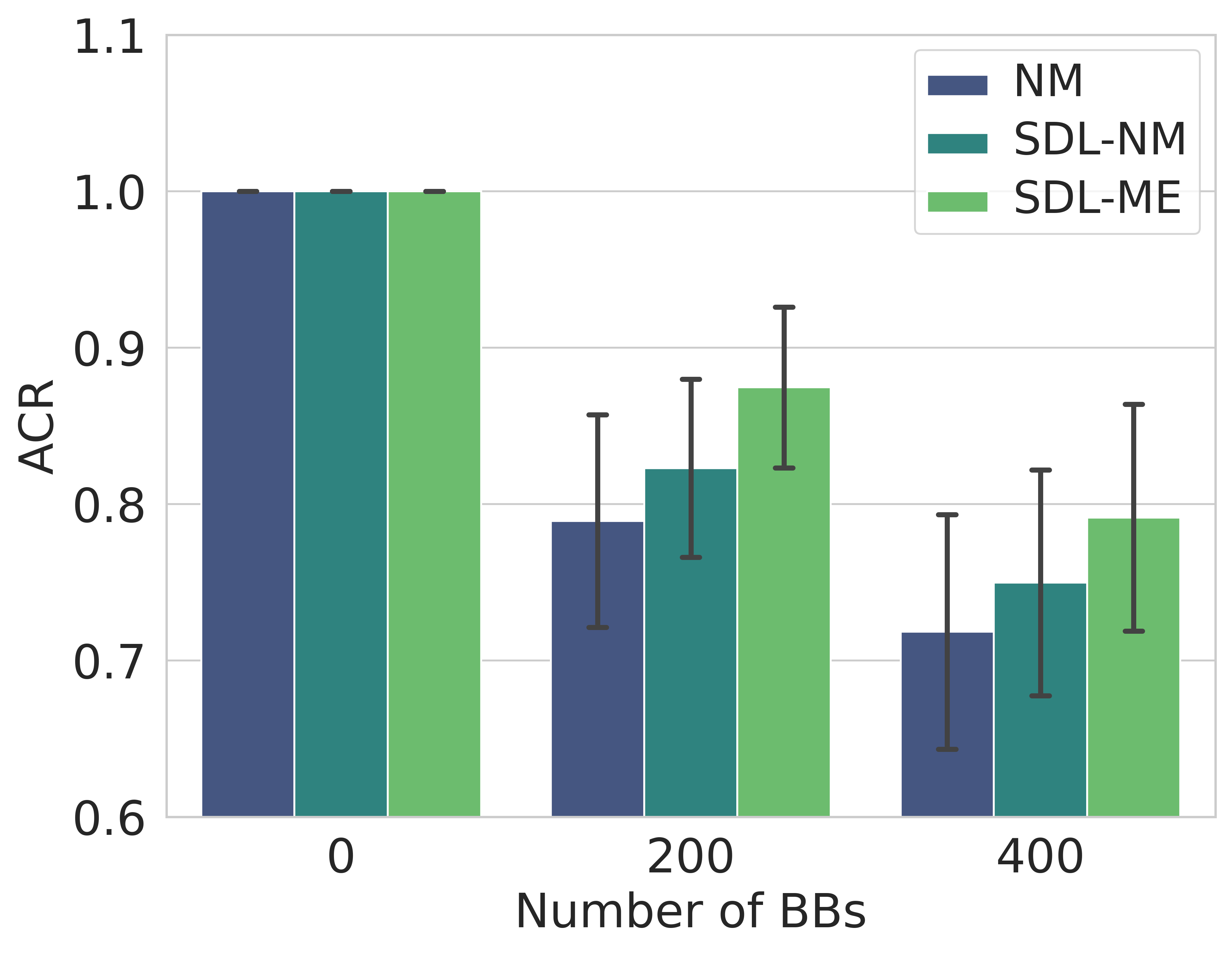}
	\caption{\rev{ACR performance of NM, SDL-NM, and SDL-ME with a different number of BBs ($N=5$, $M=100$). The height of bars represents the average value of the performance metric, while the black error bars indicate the standard deviations. Similar representation is used in the following figures.}} 
\vspace{-2ex}
	\label{fig:case_2_1}
\end{figure}

Here we extend to the scenarios with a larger problem size (e.g., with $N=5$ ABSs, $M=100$ GUs and a target throughput of $\bar R=0.83$ Mbps per GU), and compare the schemes under \textit{different site-specific environments} (e.g., with the number of BBs from 0 to 400 in the considered area).
Note that TD3 fails to converge in such larger problems.
The ACR performance of the NM, SDL-NM and SDL-ME schemes are shown in Fig. \ref{fig:case_2_1}.
It can be seen that the ACR of the three schemes decreases in general as the number of BBs increases, due to more blockages and more complex propagation environment. On the other hand, the SDL-ME approach consistently outperforms SDL-NM and NM in all the tested scenarios.

In addition, for a larger problem (e.g., with $N=5$, $M=100$), we also compare the ACR performance and robustness of NM, SDL-NM, and SDL-ME under \textit{different GU speeds}, whose model parameters are all offline trained under the default GU speed (i.e., $V_{q}=2$ m/s).
Similar to the observations in Fig. \ref{fig:case_1_2}, it can be seen that SDL-ME achieves consistently higher ACR than those of other schemes under all the tested GU speeds.
Moreover, even when tested under a different GU speed other than the one used for training, our proposed SDL-based methods suffer only minor performance degradation under small/moderate GU speeds, due to similar reasons discussed for Fig. \ref{fig:case_1_2}.
Therefore, the above results demonstrate the easy accommodation of our proposed methods to larger problems and/or with more complex/dynamic scenarios. 

\begin{figure}[!t]
	\centering
	\includegraphics[width=0.42\textwidth]{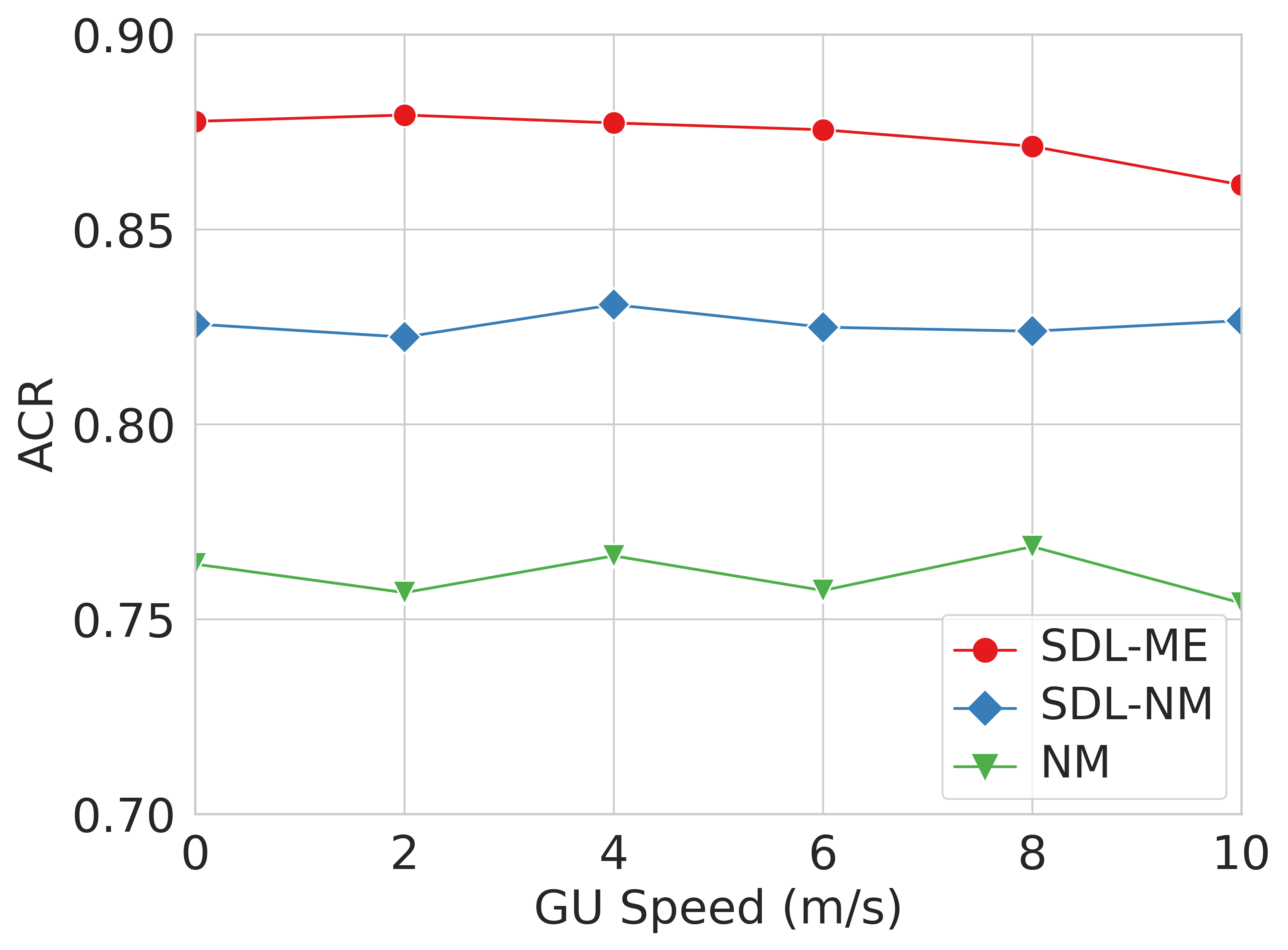}
	\caption{\rev{ACR performance of NM, SDL-NM, and SDL-ME under different GU moving speeds ($N=5$, $M=100$). The model parameters of each scheme are offline trained under $V_{q}=2$ m/s.}}\vspace{-2ex}
	\label{fig:case_2_2}
\end{figure}


\subsection{Working Mechanisms behind SDL-ME}
In this subsection, we illustrate the working mechanisms behind SDL-ME using the default settings, e.g., with $N=5$, $M=100$, and $V_{q}=2$m/s. The effectiveness of SDL-based emulator and Map-Elites based quality-diversity search is demonstrated, respectively.

\subsubsection{SDL-based Emulator for Coverage Prediction}
The SDL-based emulator takes the ABS and GU patterns as input, and predicts the CGU pattern and hence the resulting CR. To verify the accuracy of such prediction, for each period, we compare the predicted CR in emulator against the actual CR in environment for $N_\textrm{m}$ searched ABS patterns.\footnote{Note that the performance evaluation in environment for all $N_\textrm{m}$ patterns is performed here only for validation purpose. In practice, our proposed PES scheme only performs a small number of on-site performance evaluation in environment.}
Instead of predicting exactly the best ABS pattern out of the $N_\textrm{m}$ searched patterns, we target at finding the \textit{set} of top $k$ best performing ABS patterns, irrespective of the actual performance ranking order among them.
To this end, we define the \textit{successful prediction probability (SPP)} of top-$k$ patterns, denoted by $p_{k}$, as the number of actual top-$k$ candidate solutions recommended by the emulator divided by $k$.\footnote{For example, suppose $k=4$ and there are $N_\textrm{m}=100$ different ABS patterns generated during the planning phase. If the actual performance ranking order of top-$k$ predicted patterns are $1$, $3$, $2$ and $6$, then we have $p_{k}=0.75$ since one pattern is out of range. On the other hand, if the actual performance order of top-$k$ predicted patterns are $2$, $1$, $3$ and $4$, then we have $p_{k}=1$.}
Such a metric relaxes the requirement of predicting the exact performance ranking order of candidate solutions, thus reducing the accuracy requirement of the emulator and also bringing more robustness for practical implementation.
The SPP results of SDL-NM and SDL-ME averaged over $100$ independent periods are summarized in Table \ref{tab:topk}. 
It can be seen that our emulator is qualified as an environment surrogate for finding the top candidate solutions among the generated subset of solution space. In particular, with $k=10$, the SPP is close to 1 for both SDL-NM and SDL-ME.

\begin{table}[htbp]\footnotesize
	\centering
	\caption{\rev{SPP of hitting top-$k$ patterns by emulator predictions.}}
	\label{tab:topk}
 \addtolength{\tabcolsep}{-3pt}
		\begin{tabular}{|c|c|c|c|c|c|c|c|c|c|c|}
			\hline
			\diagbox{$p_{k}$}{$k$}  & 1 & 2 & 3 & 4 & 5 &6&7&8&9&10 \\ \hline
			\textbf{SDL-NM} & 0.13 & 0.30 & 0.51 & 0.67 &0.80 &0.84&0.89&0.93&0.97&0.98   \\ \hline
			\textbf{SDL-ME}  & 0.27 & 0.57 & 0.83 & 0.93 & 0.97 &0.99&1.0&1.0&1.0&1.0             \\ \hline
		\end{tabular}%
\end{table}


\subsubsection{MAP-Elites as Quality-Diversity Search Engine}
\begin{figure*}[!t]
	\centering
	\begin{subfigure}[b]{0.32\textwidth}
		\centering
		\includegraphics[width=\textwidth]{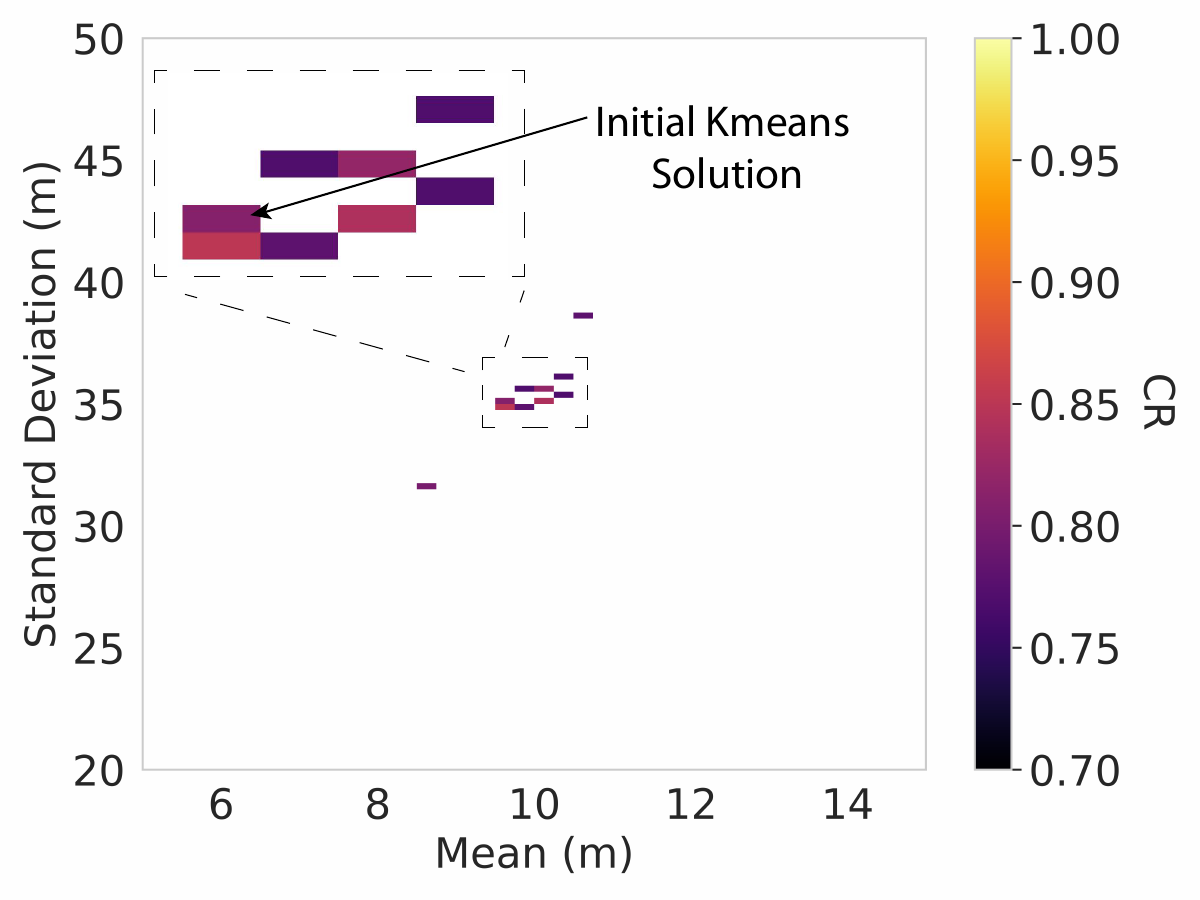}
		\caption[NM]%
		{{\small NM}}   \vspace{-1ex} 
		\label{fig:heatmap_NM}
	\end{subfigure}
	\hfill
	\begin{subfigure}[b]{0.32\textwidth}  
		\centering 
		\includegraphics[width=\textwidth]{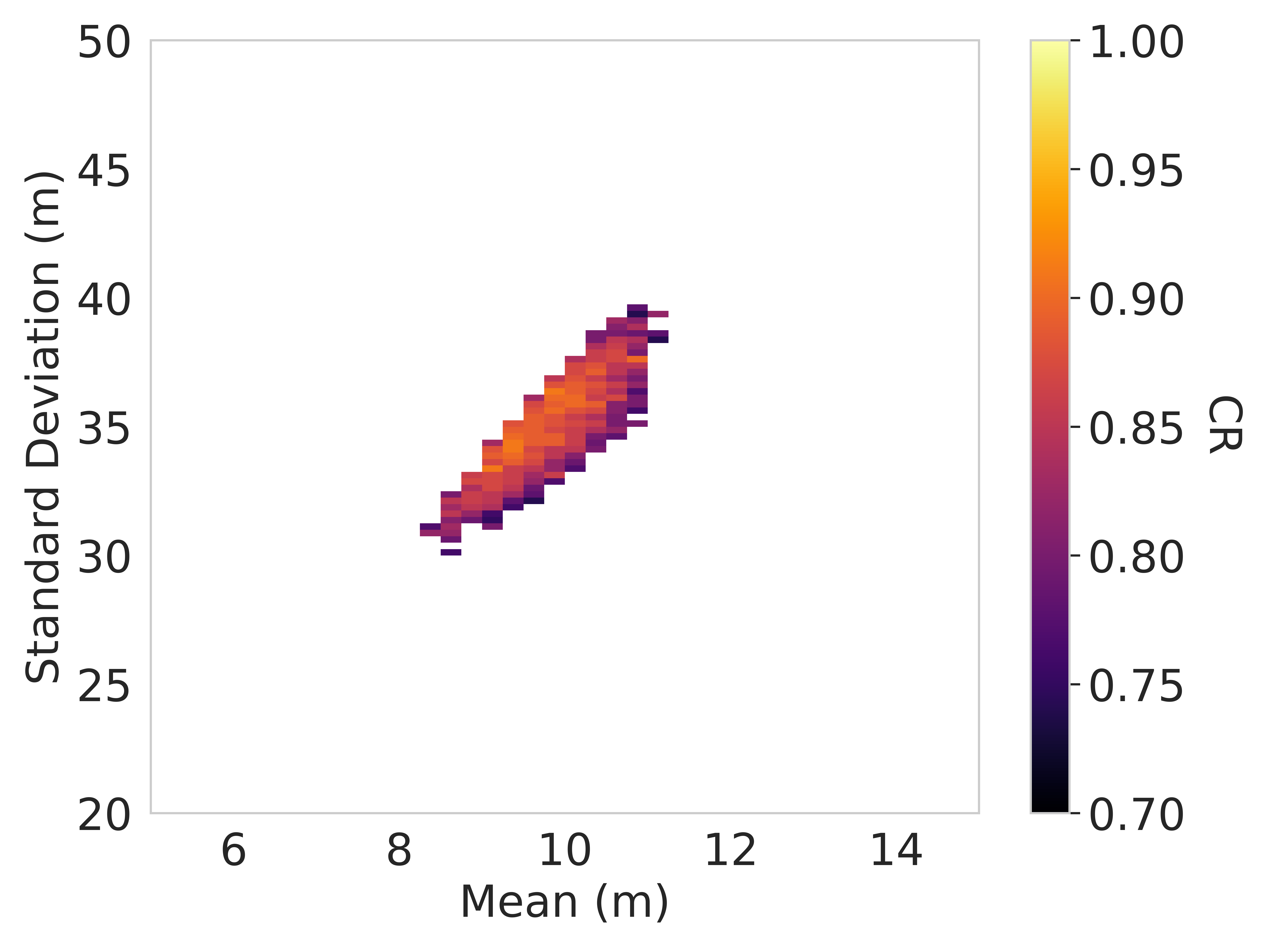}
		\caption[NM]%
		{{\small SDL-NM}}    \vspace{-1ex}
		\label{fig:heatmap_SDL_NM}
	\end{subfigure}
	\hfill
	\begin{subfigure}[b]{0.32\textwidth}   
		\centering 
		\includegraphics[width=\textwidth]{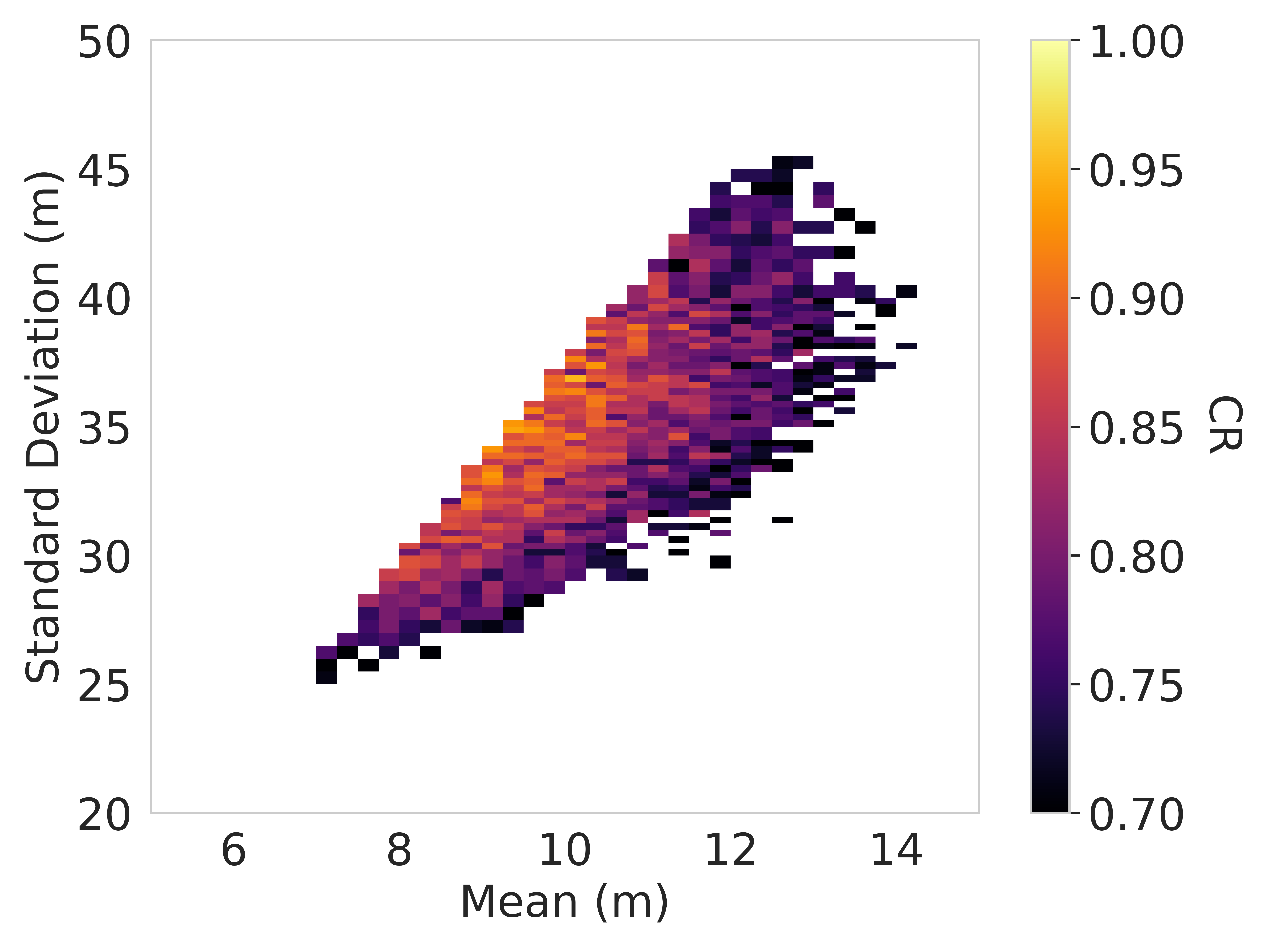}
		\caption[SDL-ME]%
		{{\small SDL-ME}}    \vspace{-1ex}
		\label{fig:heatmap_SDL_ME}
	\end{subfigure}
	\caption{\rev{Actual CR performance heatmaps in the feature space for the candidate ABS patterns ($N=5$, $M=100$).}}\vspace{-2ex}
	\label{fig:heatmap}
\end{figure*}

For a given period, we illustrate the \textit{actual CR performance heatmaps} of candidate solutions generated by NM, SDL-NM and SDL-ME methods in our selected feature space, as shown in Fig. \ref{fig:heatmap}, whose two feature dimensions are chosen as the mean and standard deviation of the Euclidean distances between ABS pairs, respectively.
The brightness of each data point (or feature niche) indicates the candidate's actual CR performance in the site measurement.
\rev{Since the NM method does not employ the emulator, there are only a few feature points around that of the initial K-means solution, and hence the coverage rates are relatively low as in Fig. \ref{fig:heatmap_NM}.}
In contrast, based on the emulator, SDL-NM and SDL-ME can virtually test a large amount of ABS patterns, thus yielding more candidate solutions in diverse feature niches, as shown in
Fig. \ref{fig:heatmap_SDL_NM} and Fig. \ref{fig:heatmap_SDL_ME}.
In particular, SDL-ME is designed to delicately mutate and search in the chosen feature space, whereby the found candidate solutions are hence more diverse, and ultimately better in general.

In summary, our proposed PES scheme seamlessly amalgamates the emulator-based planning and the on-site exploration/serving, whereby 1) the top $k$ candidate solutions are \textit{sifted out with high probability}, thus significantly reducing the need for extensive step-by-step on-site trials and errors; 2) on-site measurements for the top candidate solutions effectively compensate for the quantization/prediction errors due to model approximation; and 3) SDL-ME is designed to delicately balance between the \textit{quality and diversity} of solutions in the feature niches, which encourages more efficient search for globally better solutions.

\subsection{Practical Considerations and Robustness of SDL-ME}
In this part, we first discuss the online planning time for each period, evaluate the ACR sensitivity against the grid resolution $K$, and then demonstrate the robustness of our proposed SDL-ME method against a variable number of ABSs/GUs for on-site deployment and adaptation.



\subsubsection{Planning Time for Each Period}
The online planning is performed before the start of each period, whose running time is generally required to be comparable with or even smaller than the duration of a time period, so that the movement control can timely adapt to UE location updates. 
The planning time is mainly determined by the maximum number of mutations $N_{\text{m}}$ and the computational power of the planning agent. 
Under the default parameter settings, e.g., with time period $\Delta t=10$ s and $N_{\text{m}}=8192$, the planning time on our computer is summarized in Table \ref{tab:runtime_consumption} for different schemes with small or larger problem size.
The SDL-based methods take longer planning time compared with the baseline NM method due to emulator-enquiring operations. 
Nevertheless, the planning time of both SDL-NM and SDL-ME is controlled within 3 s in the considered setup,\footnote{The planning time can be further reduced by employing parallel/batch operations, which is left as future work.} and hence can make timely adaptation for moderately long time period (e.g., 10 s) while achieving desired performance improvements as demonstrated in the above subsections.

\begin{table}[htbp]\small
\centering
\caption{Average planning time per period for NM, SDL-NM and SDL-ME with small or larger problem size.}\label{tab:runtime_consumption}
\resizebox{0.9\linewidth}{!}{%
\begin{tabular}{|c|c|c|c|}
\hline
\textbf{Planning Time (s)}      & \textbf{NM} & \textbf{SDL-NM} & \textbf{SDL-ME} \\ \hline
$N=2$, $M=20$  & 0.001       & 1.164           & 1.534           \\ \hline
$N=5$, $M=100$ & 0.002       & 2.822           & 2.603           \\ \hline
\end{tabular}%
}
\end{table}

\subsubsection{ACR Sensitivity against Grid Resolution $K$}
\begin{figure}[!t]
\centering
\includegraphics[width=0.85\linewidth]{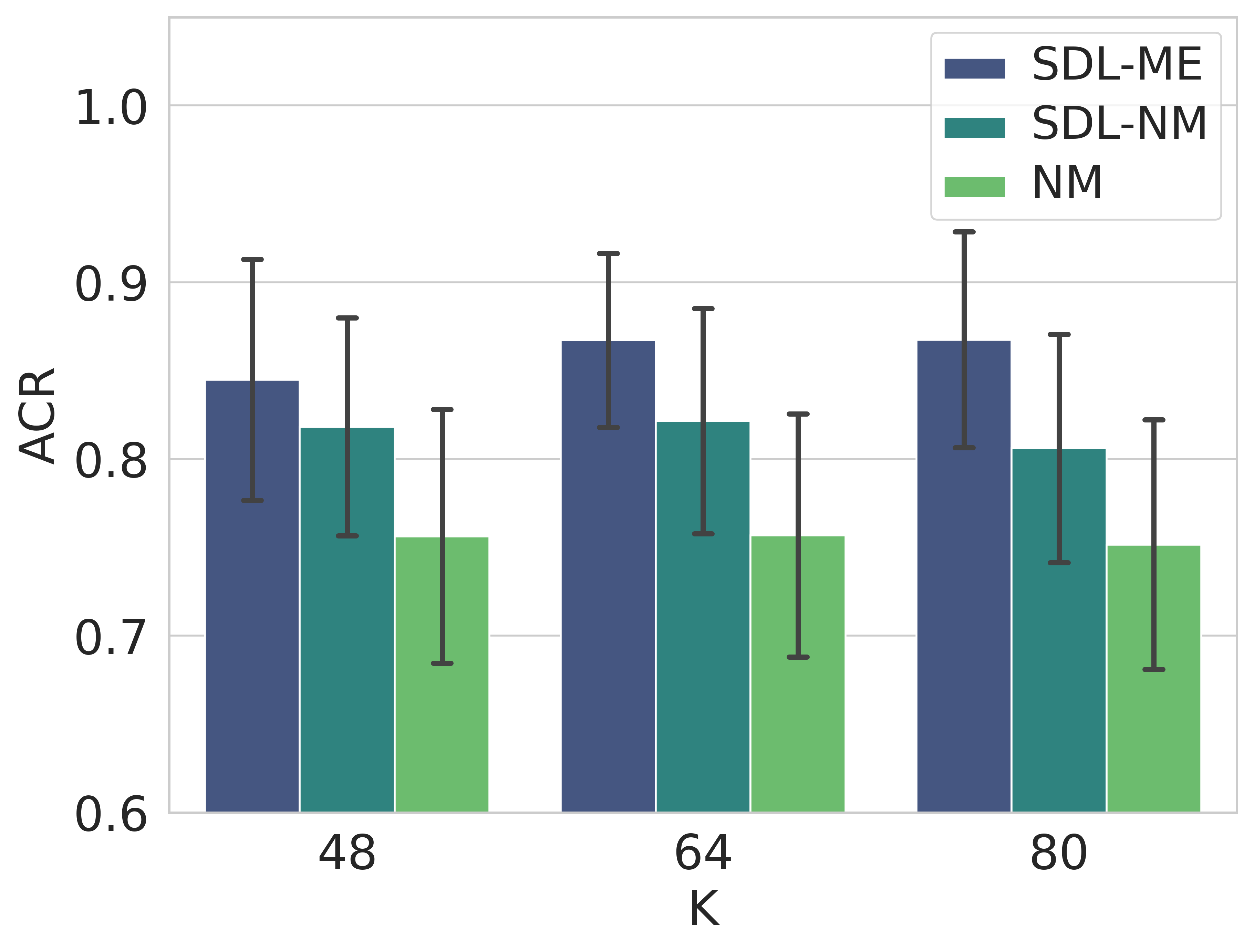}
\caption[]{\rev{ACR performance under different grid resolution $K$.}}\vspace{-2ex}
\label{case_3_K_analysis}
\end{figure}

\rev{Here we evaluate the sensitivity of the ACR performance against the grid resolution $K$ around the default value of 64 used above. It can be seen from Fig. \ref{case_3_K_analysis} that the ACR performance is relatively stable under grid resolution around $K=64$. Note that increasing the grid resolution $K$ brings more accurate quantization of ABS/GU locations at the expense of increased computational complexity, which needs to be chosen based on the applications and the available computing resources.}

\subsubsection{Robustness against a Variable Number Of ABSs/GUs}
\begin{figure}[!t]
\centering
\includegraphics[width=0.85\linewidth]{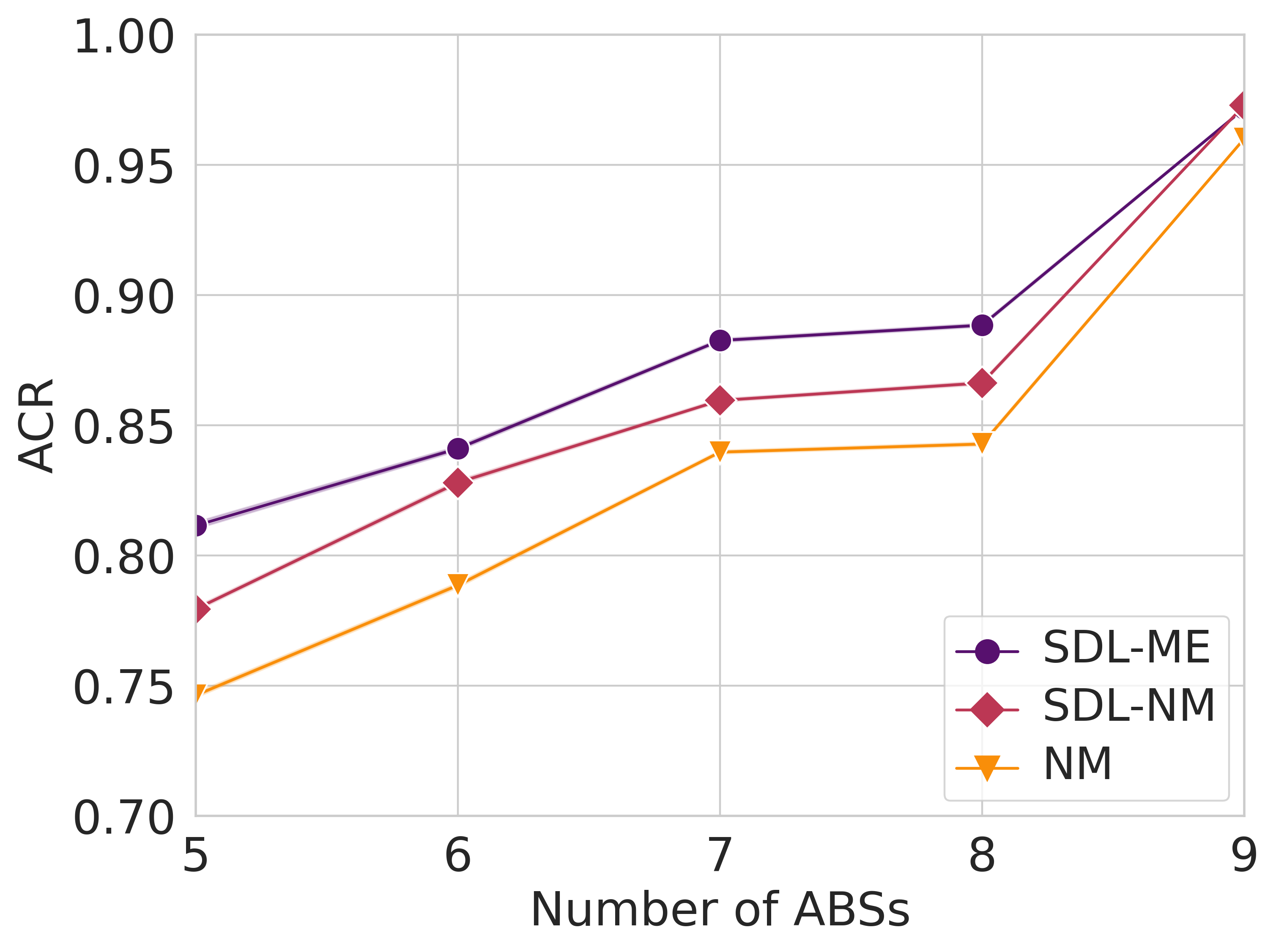}
\caption[]{\rev{ACR performance with $M=100$ and a variable number of ABSs using a single trained emulator model.}}\vspace{-2ex}
\label{fig:case_3_1}
\end{figure}

\begin{figure}[htbp]
\centering
\includegraphics[width=0.85\linewidth]{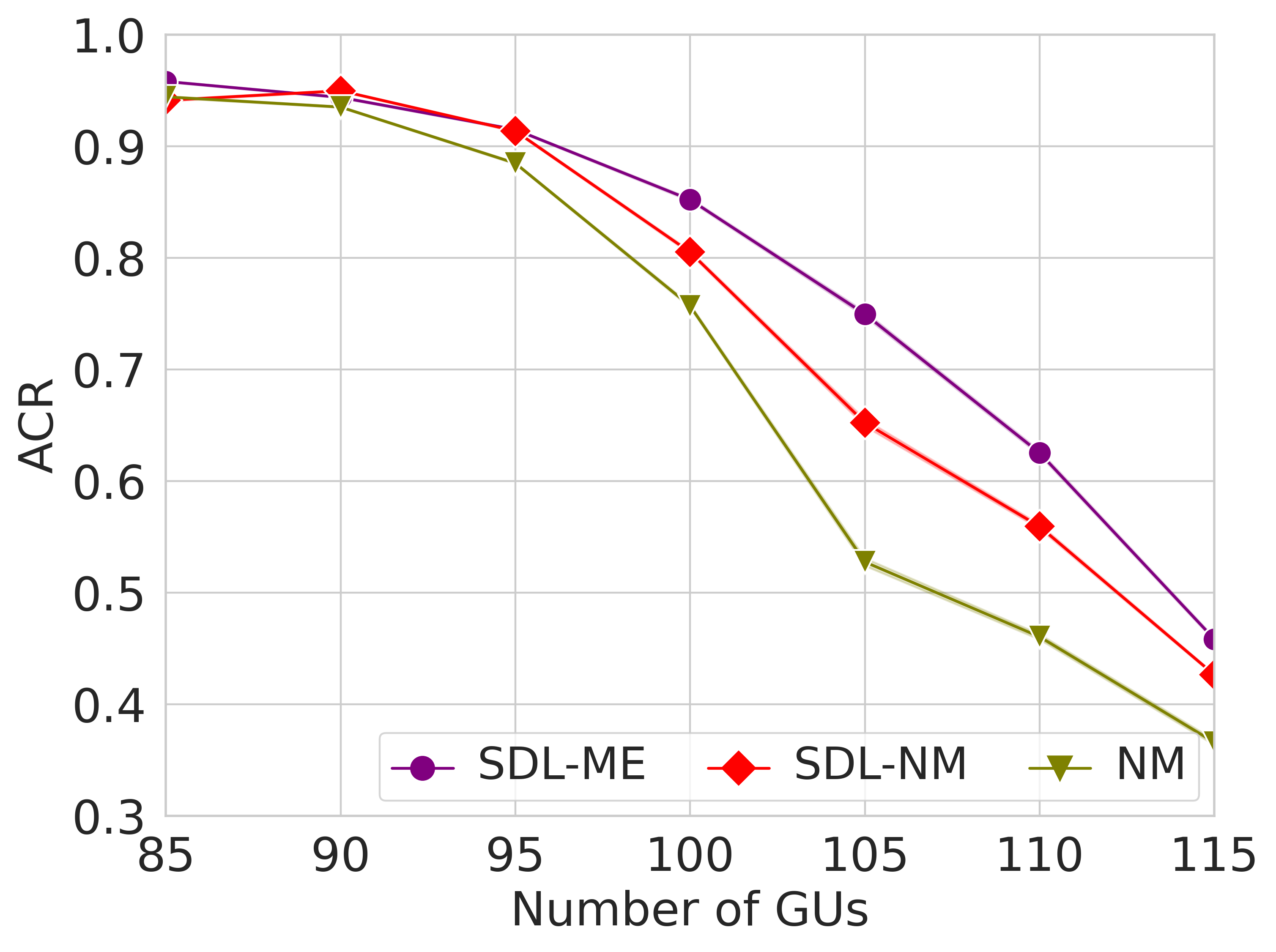}
\caption{\rev{ACR performance with $N=5$ and a variable number of GUs using a single trained emulator model.}}\vspace{-2ex}
\label{fig:case_3_2}
\end{figure}

\rev{For the practical scenarios where the number of ABSs/GUs dynamically changes on site (e.g., due to ABS malfunctioning/recharging, GU switching on/off, etc.),
our proposed SDL methods exhibit robustness against network changes without NN re-training, thanks to our grid-map based input/output design.}
Here we first demonstrate the case with an increasing number of ABSs on site, and compare the ACR performance of NM, SDL-NM and SDL-ME under a single trained emulator model, as shown in Fig. \ref{fig:case_3_1}.
It can be seen that the ACR increases in general as more ABSs become available, and SDL-ME achieves consistently better performance compared with SDL-NM and NM.
Similarly, for the case with an increasing number of GUs, the SDL methods can still be applied, whereby the ACR gradually decreases due to more GUs sharing the limited resources in the network, as shown in Fig. \ref{fig:case_3_2}.
\rev{In both cases, our SDL methods demonstrate \textit{robustness} against network changes. The underlying reason is that the proposed emulator model allows taking in diverse samples with a different number of ABSs/GUs without changing the NN input/output structures.
Therefore, by one-time training with diverse samples, the emulator model can be naturally applied in scenarios where the number of ABSs/GUs varies in a certain range.}

\section{Conclusion}\label{sec:conclusion}
This paper investigates the movement optimization of multiple ABSs to maximize the average coverage rate of mobile GUs in a site-specific environment.
The problem is NP-hard in general and further complicated by the complex propagation environment and GU mobility.
To tackle this challenging problem, a novel SDL-ME algorithm is proposed which 1) partitions the complicated ABS movement problem into ABS placement sub-problems each spanning finite time horizon; 2) designs an encoder-decoder DNN with permutation-invariant grid-map input/output structure, which serves as the environment emulator to capture spatial correlations of ABSs/GUs and thereby reduce interaction costs;
3) employs MAP-Elites as the quality-diversity search engine to efficiently search in the designated feature space (of much lower dimensions) for globally better solutions;
and 4) proposes a planning-exploration-serving scheme to seamlessly amalgamate the virtual emulator-planning with the actual site-deployment, which complement each other for reducing emulator prediction errors and speeding up on-site discovery of better solutions.
Numerical results demonstrate that the proposed approach significantly outperforms the benchmark DRL-based method and other two baselines in terms of ACR, training time and/or sample efficiency.
Moreover, with one-time training, our proposed method can be applied in scenarios where the number of ABSs/GUs dynamically changes on site and/or with different/varying GU speeds, which is thus more robust and flexible compared with conventional DRL-based methods.




{\small
	\bibliography{IEEEabrv,ref2}}

\begin{IEEEbiography}[{\includegraphics[width=1in,height=1.25in,clip,keepaspectratio]{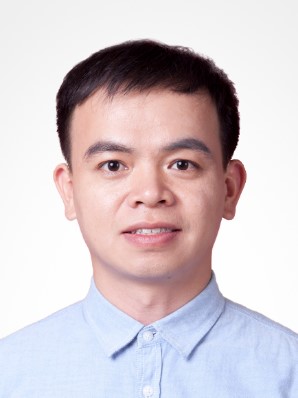}}]{Jiangbin Lyu}
	(S'12, M'16) received his B. Eng. degree (Hons.) under the Chu Kochen Honors Program from Zhejiang University, Hangzhou, China, in 2011, and the Ph.D. degree from NUS Graduate School for Integrative Sciences and Engineering (NGS), National University of Singapore (NUS), Singapore, in 2015. 
	
	He was a Post-Doctoral Research Fellow in NUS from 2015 to 2017. He is currently an associate professor in the School of Informatics, Xiamen University, China, with research interests in unmanned aerial vehicle communications, intelligent reflecting surface, radio map, etc. 
	
	Dr. Lyu was a first-author recipient of the IEEE Communications Society Heinrich Hertz Prize Paper Award in 2020, and also the Best Paper Award at Singapore-Japan International Workshop on Smart Wireless Communications in 2014. He served as the Invited Track Co-Chair at the 2021 IEEE/CIC ICCC conference, a CAA committee member of Industrial Internet of Things Technology and Applications, a TPC member for IEEE GLOBECOM and ICC, and a reviewer for various IEEE journals.	
\end{IEEEbiography}
\begin{IEEEbiography}[{\includegraphics[width=1in,height=1.25in,clip,keepaspectratio]{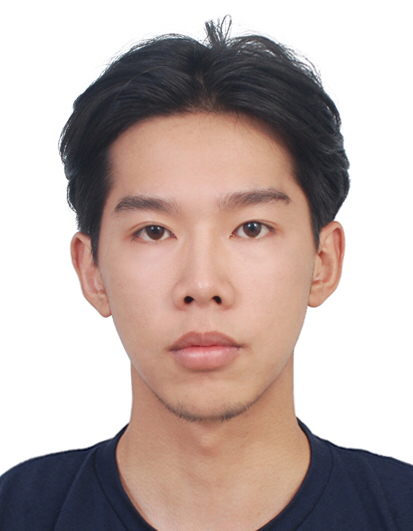}}]{Xu Chen}
	received the B.S. degree with Honours in computer science from Dalhousie University, Halifax, Canada, in 2019, and the M.S. degree in information and communication engineering from Xiamen University, Xiamen, China, in 2023. His research interests include unmanned aerial vehicle communications and deep learning.
\end{IEEEbiography}
\begin{IEEEbiography}[{\includegraphics[width=1in,height=1.25in,clip,keepaspectratio]{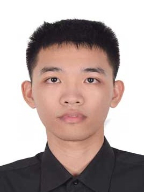}}]{Jiefeng Zhang}
	received the B.S. degree in communication engineering from Xiamen University, Xiamen, China, in 2019, where he is currently working toward the M.S. degree in information and communication engineering. His research interests include unmanned aerial vehicle communications and the application deep learning method.
\end{IEEEbiography}
\begin{IEEEbiography}[{\includegraphics[width=1in,height=1.25in,clip,keepaspectratio]{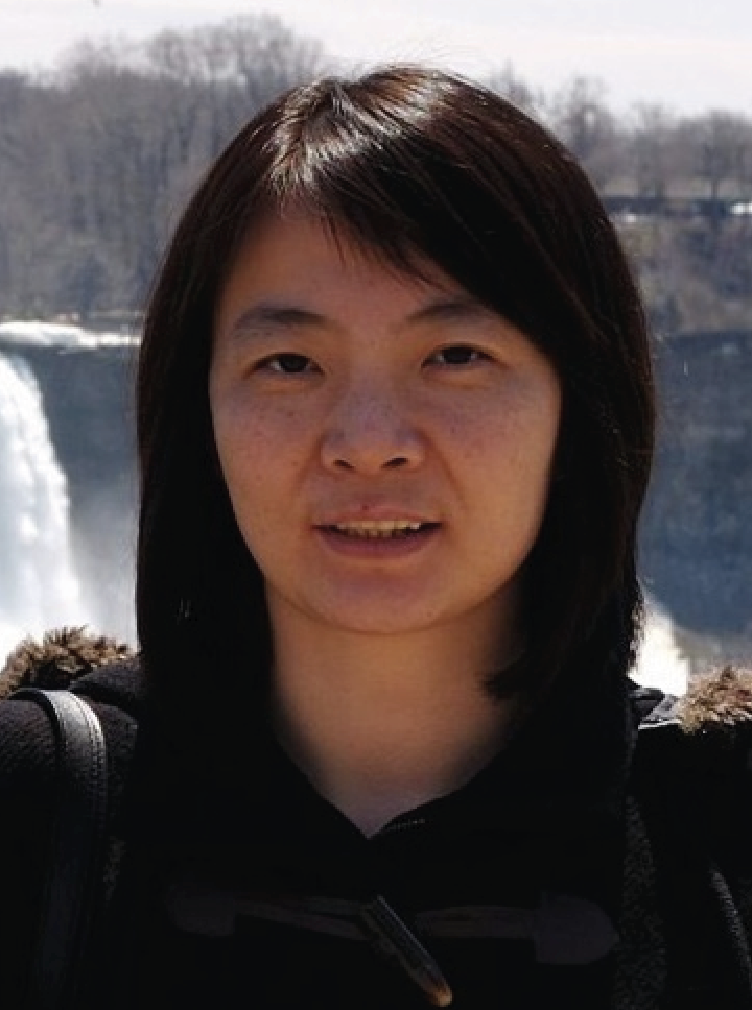}}]{Liqun Fu} (S'08-M'11-SM'17) is a Full Professor of the School of Informatics at Xiamen University, China. She received her Ph.D. Degree in Information Engineering from The Chinese University of Hong Kong in 2010. She was a post-doctoral research fellow with the Institute of Network Coding of The Chinese University of Hong Kong, and the ACCESS Linnaeus Centre of KTH Royal Institute of Technology during 2011-2013 and 2013-2015, respectively. She was with ShanghaiTech University as an Assistant Professor during 2015-2016.
	
	Her research interests are mainly in communication theory, optimization theory, game theory, and learning theory, with applications in wireless networks. She is on the editorial board of IEEE Communications Letters and the Journal of Communications and Information Networks (JCIN). She served as the Technical Program Co-Chair of IEEE/CIC ICCC 2021 and the GCCCN Workshop of the IEEE INFOCOM 2014, the Publicity Co-Chair of the GSNC Workshop of the IEEE INFOCOM 2016, and the Web Chair of the IEEE WiOpt 2018. She also serves as a TPC member for many leading conferences in communications and networking, such as the IEEE INFOCOM, ICC, and GLOBECOM.
\end{IEEEbiography}

\newpage
\ifCLASSOPTIONcaptionsoff
\newpage
\fi

\end{document}